\documentclass[journal, 12pt, onecolumn, draftcls]{IEEEtran}

\ifCLASSINFOpdf
\else
\fi
\usepackage{xcolor}
\usepackage{algorithmic}
\usepackage{array}
\usepackage{stfloats}
\usepackage{cite}
\usepackage{graphicx}
\graphicspath{{Figures/}{../Figures/}}
\usepackage{gensymb,bm}
\usepackage[T1]{fontenc}
\usepackage[utf8]{inputenc}
\usepackage{pdfpages}
\usepackage{textcomp}
\usepackage[cmex10]{amsmath}
\usepackage{booktabs}
\usepackage{url}
\usepackage{subcaption}
\usepackage{url}
\usepackage{multicol}
\usepackage{multirow}
\begin{document}

\title{THz Band Channel Measurements and Statistical Modeling for Urban Microcellular Environments}
% need to check the membership status if the DRs; maybe they aren't members
\author{Naveed A. Abbasi,
	Jorge~Gomez-Ponce, Revanth Kondaveti,
	Ashish Kumar,
	Eshan Bhagat, Rakesh N S Rao,
	Shadi Abu-Surra, Gary Xu, Charlie Zhang,
	and~Andreas~F.~Molisch% <-this % stops a space
	\thanks{The work of USC was partly supported by the Semiconductor Research Corporation (SRC) under the ComSenTer program, Samsung Research America, the National Science Foundation, the National Institute for Standards and Technology, and the Foreign Fulbright Ecuador SENESCYT Program. }
	\thanks{N. A. Abbasi, J. Gomez-Ponce, R. Kondaveti, A. Kumar, E. Bhagat, Rakesh N S Rao and A. F. Molisch are with the Ming Hsieh Department of Electrical and Computer Engineering, University of Southern California, Los Angeles, CA, USA. J. Gomez-Ponce is also with the ESPOL Polytechnic University, Escuela Superior Politécnica del Litoral, ESPOL, Facultad de Ingenier\'ia en Electricidad y Computaci\'on, Km 30.5 vía Perimetral, P. O. Box 09-01-5863, Guayaquil, Ecuador. S. Abu-Surra, G. Xu and C. Zhang are with Samsung Research America, Richardson, TX, USA. Corresponding author: Naveed A. Abbasi (nabbasi@usc.edu).}% <-this % stops a space
}% <-this % stops a space
\maketitle

\begin{abstract}
	The THz band (0.1-10 THz) has attracted considerable attention for next-generation wireless communications, due to the large amount of available bandwidth that may be key to meet the rapidly increasing data rate requirements. Before deploying a system in this band, a detailed wireless channel analysis is required as the basis for proper design and testing of system implementations. One of the most important deployment scenarios of this band is the outdoor microcellular environment, where the Transmitter (Tx) and the Receiver (Rx) have a significant height difference (typically $ \ge 10$ m). In this paper, we present double-directional (i.e., directionally resolved at both link ends) channel measurements in such a microcellular scenario encompassing street canyons and an open square. Measurements are done for a 1 GHz bandwidth between 145-146 GHz and an antenna beamwidth of 13 degree; distances between Tx and Rx are up to 85 m and the Tx is at a height of 11.5 m from the ground. The measurements are analyzed to estimate path loss, shadowing, delay spread, angular spread, and multipath component (MPC) power distribution. These results allow the development of more realistic and detailed THz channel models and system performance assessment.
\end{abstract}

\begin{IEEEkeywords}
	THz Channel Measurements, Outdoor Channel, Urban Scenario, Statistical Modeling, Microcellular
\end{IEEEkeywords}
\IEEEpeerreviewmaketitle

\section{Introduction}
A number of new and upcoming applications require ultra-high data rates that are beyond the capabilities of mmWave-based 5G communication systems. In order to meet these requirements, higher frequencies such as the THz band (0.1-10 THz) are being investigated because of the availability of considerable amounts of unused spectrum in these bands \cite{Tataria_6G,5764977,huq2019terahertz,rappaport2019wireless}. Therefore the THz band, especially the frequencies between 0.1-0.5 THz, has been explored by a number of studies, e.g., \cite{Kurner2014,6898846,khalid2019statistical,ju2021subterahertz}. The recent decision of the Federal Communication Commission (FCC), the US spectrum regulator, to provide experimental licenses in this band has fostered additional research interest, and this band is widely expected to be an important part of 6G wireless systems \cite{tataria20216g}. 

It is important to know the characteristics of a wireless channel before the design of a communication system that is to operate in it can proceed. Channel sounding measurements and their statistical analysis are  an essential  first step towards the understanding of a channel and consequently towards the design and deployment of a wireless system \cite{molisch2012wireless}. Since channel characteristics are highly dependent on the operating frequency range as well as the environment and the scenarios a wireless channel operates in, channel sounding campaigns need to be performed in the key scenarios of interest. 

Existing channel measurements in the THz bands are mostly limited to short-distance indoor channels, see \cite{priebe2010measurement,6898846,6574880,khalid2019statistical,abbasi2020channel,xing2021millimeter}, usually as a result of measurement setup constraints; see also \cite{han2021terahertz} and references therein. However, recently there has been some progress on longer distances and outdoor scenarios as well. These include the first long-distance (100 m) double-directional channel measurements for the 140 GHz band, which were reported in 2019 \cite{abbasi2019double,abbasi2020double} by our group, as well as our recent works \cite{abbasi2021ultra,abbasi2021double, Abbasi2021THz} where we target device-to-device (D2D) scenarios, where both Tx and Rx are at about 1.6 m height. Another recent series of papers \cite{xing2021propagation,ju2021subterahertz,9558848} also reported channel measurements, path loss and statistical modeling at 140 GHz over longer channel lengths in an urban scenario; in those measurements the Tx is placed at 4 m above the ground (i.e., typical lamppost height). Our current paper aims to provide analysis for a scenario where the Tx is significantly higher, at 11.5 m, which is comparable to the height of a typical microcell base station height. This paper presents the results of an extensive measurement campaign in this environment, with sufficient points to allow a meaningful statistical evaluation. To the best of our knowledge, such a detailed channel measurement campaign for cases where Tx is elevated more than 10 m above the ground has not been reported before in the THz band.

The results of this paper are based on ultra-wideband double-directional channel measurements for a 1 GHz bandwidth between 145-146 GHz\footnote{Some authors prefer to use the term "THz" to identify the frequency range $>300$ GHz while using "high mmWave", "sub-THz" or `low-THz' for frequencies between 100-300 GHz. Other authors use the term "THz" for both these cases. Since the latter is the most widely used terminology, we will employ it in this paper as well}, conducted at 26 different transmitter (Tx) - receiver (Rx) location pairs. 13 of these represent line-of-sight (LoS) scenarios with direct Tx-Rx distances ranging from nearly 20 m to 83 m, while the other 13 are non-line-of-sight (NLoS) cases with direct Tx-Rx distances also in approximately the same range. Based on the nearly 110,000 directional impulse responses we collected from these measurements, we model the path loss, shadowing, delay spread, angular spread and multipath (MPC) power distribution for both LoS and NLoS cases. Our detailed analysis includes results both for the maximum-power-beam direction (max-dir) and the omni-directional characteristics as well as the distance dependence of the key parameters, and their relevant confidence intervals for the various model fits. 

The remainder of this paper is organized as follows. In Section II, we describe the channel sounding setup and the measurement locations. Key parameters of interest and their processing is described in Section III. The results of the measurements and modeling are presented in Section IV. We finally conclude the manuscript in Section V.

\section{Measurement equipment and site}
\subsection{Testbed description}
\begin{figure}[t!]
	\centering
	\includegraphics[width=12cm]{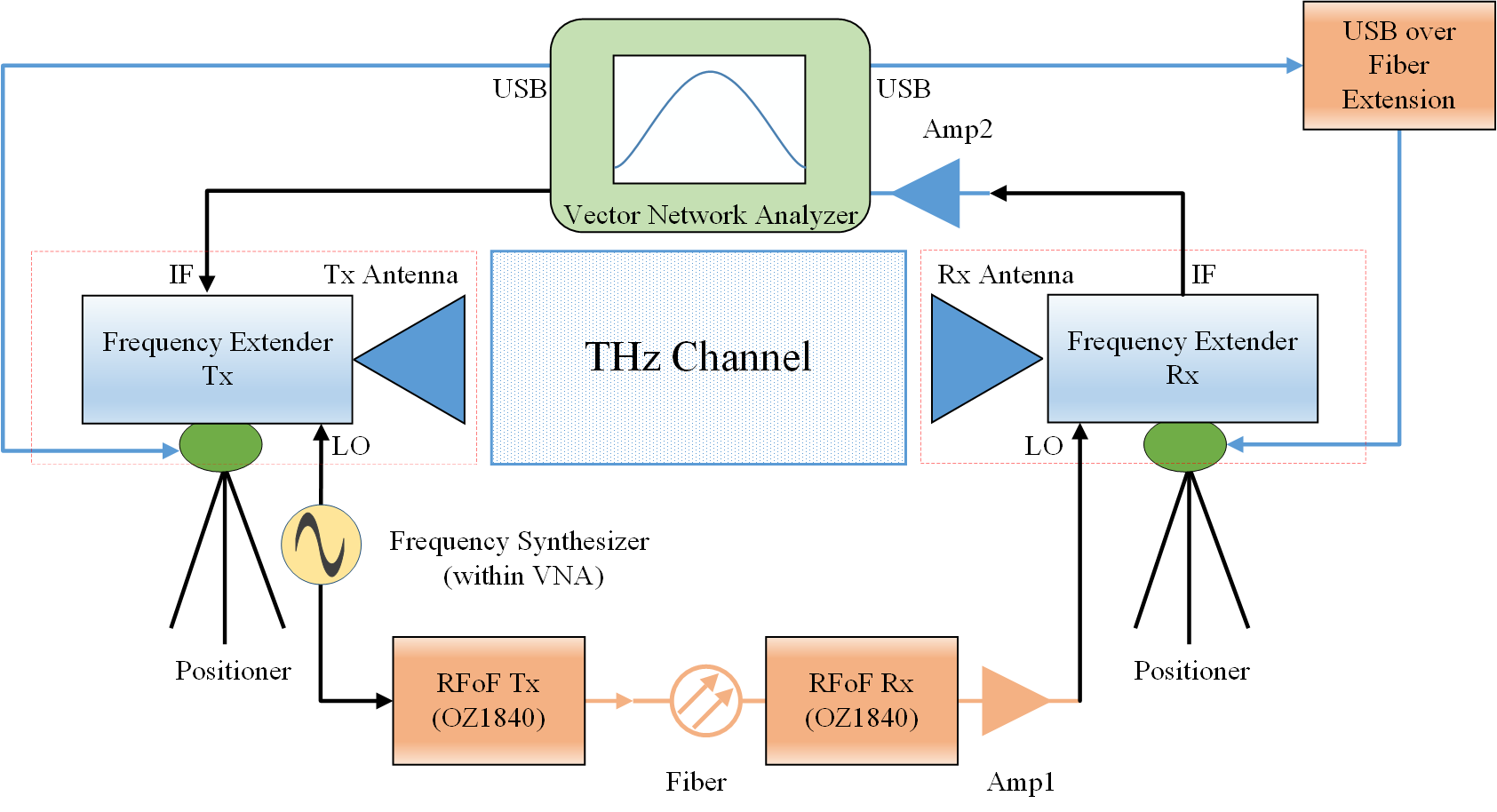}
	\caption{Channel sounding setup.}
	\label{fig:setup}
\end{figure}

\begin{table}[t!]
	%\vspace{-0.15cm}
	\centering
	\caption{Setup parameters.}
	\label{table:parameters}
	\begin{tabular}{|l|l|l|}
		\hline
		\textbf{Parameter}              & \textbf{Symbol}   & \textbf{Value} \\ \hline\hline
		\textit{Measurement points}     & $N$                 & 1001           \\
		\textit{Tx height}     & $h_{Tx}$                 & 11.5 m           \\
		\textit{Rx height}     & $h_{Rx}$                 & 1.7 m           \\
		\textit{Start frequency}        & $ f_{start} $     & 145 GHz          \\
		\textit{Stop frequency}         & $ f_{stop} $      & 146 GHz         \\
		\textit{Bandwidth}              & $BW$              & 1 GHz         \\	
		\textit{IF bandwidth}              & $IF_{BW}$              & 10 KHz         \\				
		\textit{THz IF}                   & $ f_{THz IF} $ & 279 MHz          \\
		\textit{Antenna 3 dB beamwidth}   & $\theta_{3dB}$        & 13$^{\circ}$ \\
		\textit{Tx Az rotation range}   & $\phi_{Tx}$        & [-60$^{\circ}$,60$^{\circ}$]           \\ 
		\textit{Tx Az rotation resolution}   & $\Delta \phi_{Tx}$        & 10$^{\circ}$           \\
		\textit{Rx Az rotation range}   & $\phi_{Rx}$        & [0$^{\circ}$,360$^{\circ}$]           \\
		\textit{Rx Az rotation resolution}   & $\Delta \phi_{Rx}$        & 10$^{\circ}$           \\
		\textit{Tx El rotation range}   & $\tilde{\theta}_{Tx}$        & [-13$^{\circ}$,13$^{\circ}$]           \\
		\textit{Tx El rotation resolution}   & $\Delta \tilde{\theta}_{Tx}$        & 13$^{\circ}$           \\
		\textit{Rx El rotation range}   & $\tilde{\theta}_{Rx}$        & [-13$^{\circ}$,13$^{\circ}$]           \\
		\textit{Rx El rotation resolution}   & $\Delta \tilde{\theta}_{Rx}$        & 13$^{\circ}$           \\ \hline
	\end{tabular}
\end{table}

For this measurement campaign, a frequency-domain channel sounder was used (see in Fig. \ref{fig:setup}), similar to \cite{Abbasi2021THz}. It is based on a Vector Network Analyzer (VNA), PNAX N5247A from Keysight, which has a frequency range from 10 MHz to 67 GHz. Frequency extenders, WR-5.1 VNAX manufactured by Virginia Diodes, were used to increase the VNA's frequency range to the 140-220 GHz band, which encompasses the band of interest to us. The extenders were used with the "high sensitivity" waveguide option to improve the received Signal to Noise Ratio (SNR). The antennas (along with the extenders) are mounted on a rotating positioning system. A key aspect of this setup is the use of a RF-over-fiber (RFoF) link, which was originally introduced in \cite{abbasi2020double}. The RFoF allows us to measure over longer distances than the typical 5-10 m range of similar systems without the link. For further details of the system please see \cite{Abbasi2021THz}.

Table \ref{table:parameters} shows the configuration parameters for the sounder. The IF bandwidth of the VNA was selected such that there is a compromise between the dynamic range and the measurement duration, such that the duration of a measurement sweep is lower than the mechanical movement of the horn, and therefore has only a minor impact on the total measurement time. Each sweep of the VNA contains 1001 frequency points over the 1 GHz bandwidth, therefore allowing a maximum excess delay of 1 $\mu s$ without suffering the effects of aliasing. In other words, the maximum measurable excess runlength for multipaths is 300 m, a reasonable distance considering the scenarios and the frequency band being sounded.  Given that the measurements take a significant amount of time, they were conducted at night while ensuring the scenario remains static/quasi static.

The measurement locations were selected to be typical of a "microcellular" scenario. The Tx for the current measurements is set at a height of 11.5 m above the ground while the Rx is placed 1.7 m high from the ground. These parameters have been selected following the 3GPP UMi Street Canyon model, (3GPP TR 38.901 version 14.0.0 Release 14 suggests $h_{Tx}=10m$ and $1.5m \leq h_{Rx} \leq 22.5m$). Additionally, to extract the double-directional characteristics of the channel, the frequency sweeps of the VNA were repeated with sets of different orientations of the antennas. The positioners were oriented to ensure that the azimuth angle zero at both ends (Tx and Rx) corresponded to the LoS direction, irrespective of whether an unblocked optical LoS connection between Tx and Rx actually exists or not. We anticipated that multiple elevation scans are required to properly analyze the scenario, due to the different heights of the link ends, therefore, three elevation cuts are scanned on both the Tx and Rx. The Tx azimuth will scan a $120^\circ$ sector from $-60^\circ$ to $60^\circ$ with $10^\circ$ of azimuthal resolution, meanwhile, the Rx will carry out a complete azimuth scan, from $0^\circ$ to $360^\circ$ in steps of $10^\circ$, similar to Tx. In elevation, Tx and Rx are aligned so that when both antennas are facing ($\tilde{\theta}_{Tx}=\tilde{\theta}_{Rx}=0^\circ$), they are in the same elevation cut. After that, both ends will make additional scans $13^\circ$ above and $13^\circ$ below the "alignment", giving a total of 9 elevation scans per Tx-Rx location (3 elevation scans at the Tx and 3 for the Rx). 

The measurements were performed on different days, due to the long measurement time per point. For each day a calibration of the VNA, as well an over-the-air calibration (OTA) with the Tx and Rx at a LoS location was performed. Additional details of the setup are described in \cite{abbasi2020double,abbasi2021double,abbasi2021ultra} \footnote{It is important to mention that $\tilde{\theta} = 0^\circ$ is not equivalent to $\theta = 90^\circ$ in elevation, i.e. it is not the horizontal. $\tilde{\theta} = 0^\circ$ is different on each point in an absolute elevation reference.}.\\
Finally, the frequency domain-sounder provides a high phase stability which allows to conduct Fourier analysis and High Resolution Parameter Extraction (HRPE). Although HRPE can provide more accurate results, the current paper only uses Fourier analysis; HRPE analysis will be discussed in future work. 

\subsection{Measurement locations}
\begin{figure}[ht]
	\centering
	%\hspace{7mm}
	\includegraphics[width=1\columnwidth]{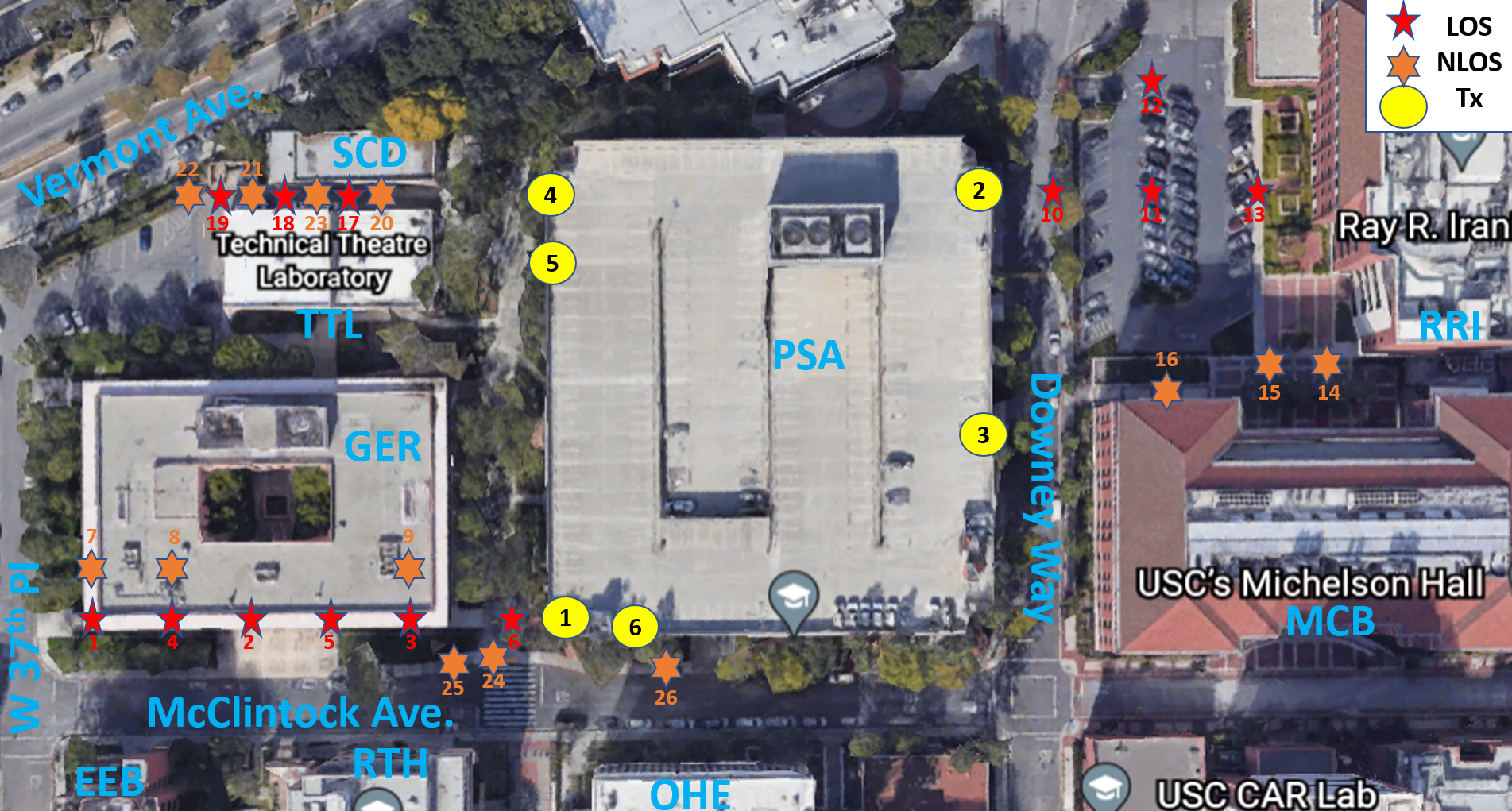}
	\caption{Microcellular campaign measurement scenario.}
	\vspace*{0mm}
	\label{fig:Micro_sce}
\end{figure}

A very important step in the measurement campaign is the selection of suitable locations so that we can realistically measure samples of LoS and NLoS scenarios. For this purpose we selected an area inside the University Park Campus of the University of Southern California (USC) in Los Angeles California, USA, that is located in the center of the city and is characterized as an urban environment. Fig. \ref{fig:Micro_sce} shows the scenario and locations of the Tx and Rx locations. As can be seen, the measurement campaign is divided into 6 routes with LoS or NLoS points each corresponding to a unique Tx location. For all 6 Tx locations, the positioner was placed on the edge of the Downey Way Parking Structure (PSA) building on the third floor. 

\begin{figure*}[t!]
	\centering
	\begin{subfigure}{0.31\textwidth}
		\centering        
		\centering
		\hspace{7mm}
		\includegraphics[width=1\columnwidth]{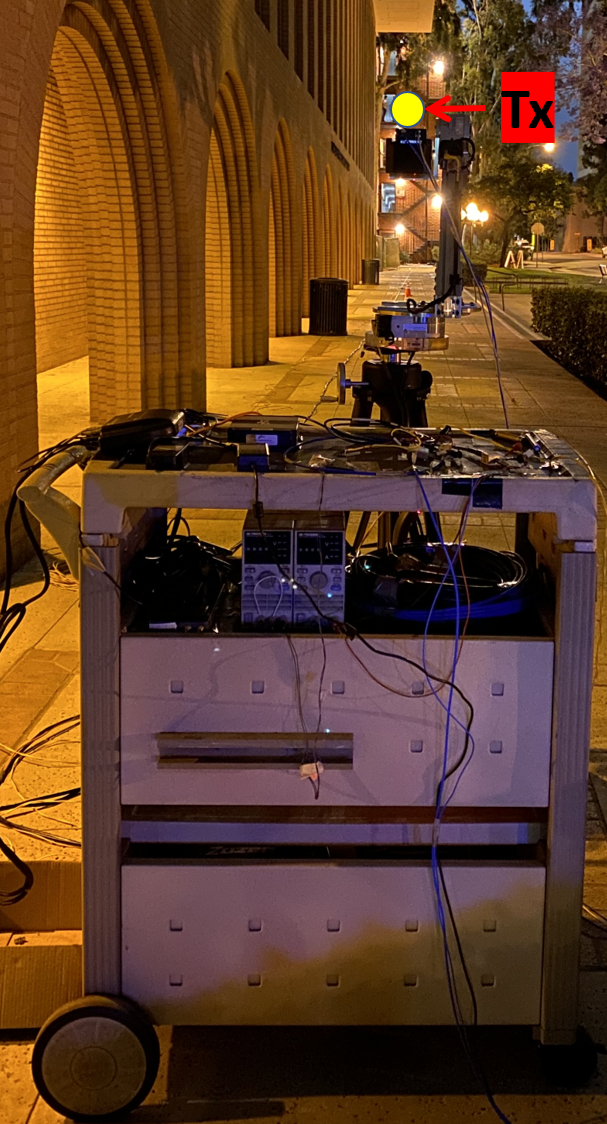}
		\caption{Tx1-Rx1 LoS; $d=81.7 m$.}
		\vspace*{0mm}
		\label{fig:LOS TX1-RX1}
	\end{subfigure}
	\begin{subfigure}{0.45\textwidth}
		\centering
		\centering
		\hspace{7mm}
		\includegraphics[width=1\columnwidth]{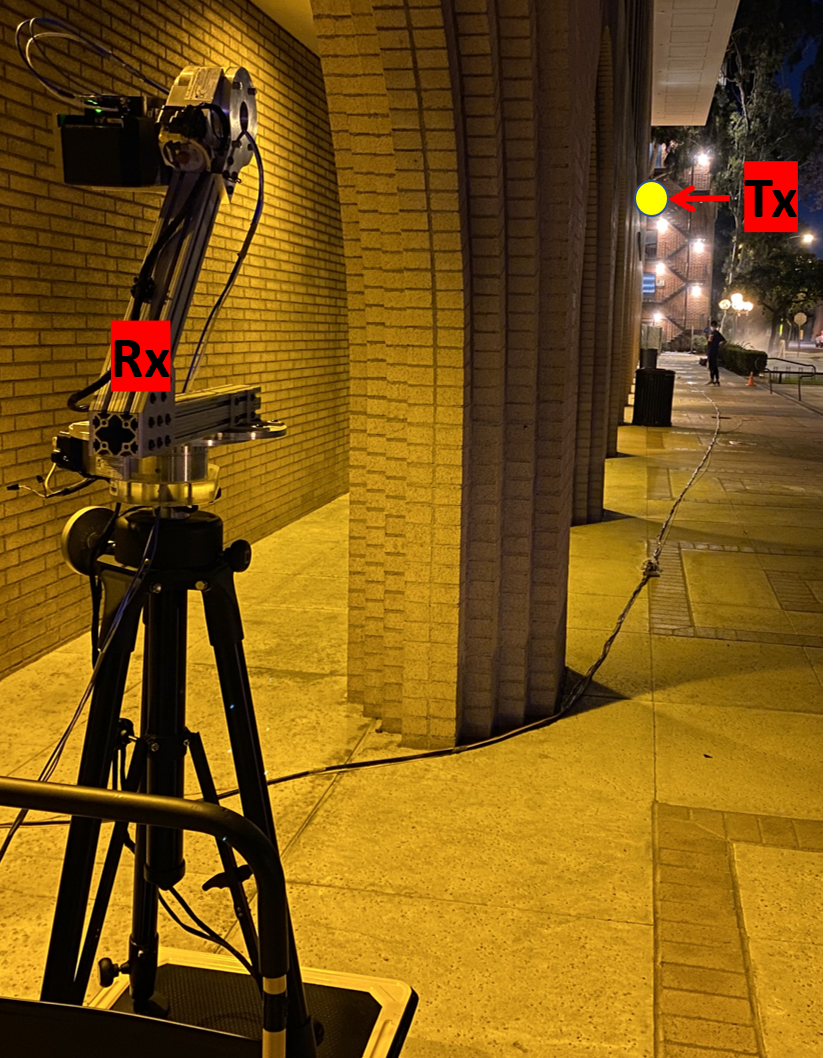} 
		\caption{Tx1-Rx7 NLoS ; $d=83.2 m$.}
		\vspace*{0mm}
		\label{fig:NLOS TX1-RX7}
	\end{subfigure}
	\caption{LoS and NLoS measurement points for Route One.}
	\label{fig:TX1}
\end{figure*}

Route One contains 6 LoS points aligned on the walkway of the Andrus Gerontology Center (GER) on the McClintock side of the building, covering a distance range from 33.5 to 81.7 m (see Fig. \ref{fig:LOS TX1-RX1}). Ronald Tutor Hall (RTH) and the Hughes Aircraft Electrical Engineering Center (EEB) together with the GER building create a "street canyon" for Route One points. It is important to note that the LoS was not obstructed or partially obstructed by foliage or other environmental objects. The three NLoS points were placed under the portico of the GER building (see Fig. \ref{fig:NLOS TX1-RX7}). Apart from the roof of the building, the pillars provide additional obstructions to the LoS. The second route is at the opposite side of PSA on a parking lot surrounded by Ray Irani (RRI) and Michelson Hall (MCB). While photo of Fig. \ref{fig:Micro_sce} shows cars, no cars were present during the measurement. Rx points 10, 11 and 13 were set on a straight line aligned to the Tx and 12 was set 30 meters north of point 11. For Route Three, the Tx is moved 40 meters along PSA parallel to Downey Way. Here, MCB's side corner completely blocks the LoS components for points 14-16. The distances for this route are approximately in the range of 40 to 60 meters. 

Route Four places the Tx in the north west corner of PSA, the three Rx locations are placed in an alley between Technical Theatre Laboratory (TTL) and the Scene Dock Theatre (SCD) buildings at distances ranging from 35 to 65 meters approximately. Route Five places the Tx 15 meters south of the Tx location in Route Four and the four Rx locations were placed in the same alley between SCD and TTL as Route Four. The obstruction for this route is provided by the TTL building and foliage as shown in Fig. \ref{fig:Micro_sce} \footnote{Delay domain results for the subset of measurements on Route Four and Five will be presented in \cite{abbasi2022double}. This analysis is significantly different from the statistical analysis of the current work, which is based on a large set of measurements.}.

\begin{figure*}[t!]
	\centering
	\begin{subfigure}{0.3\textwidth}
		\centering
		\hspace{0mm}
		\includegraphics[width=1\columnwidth]{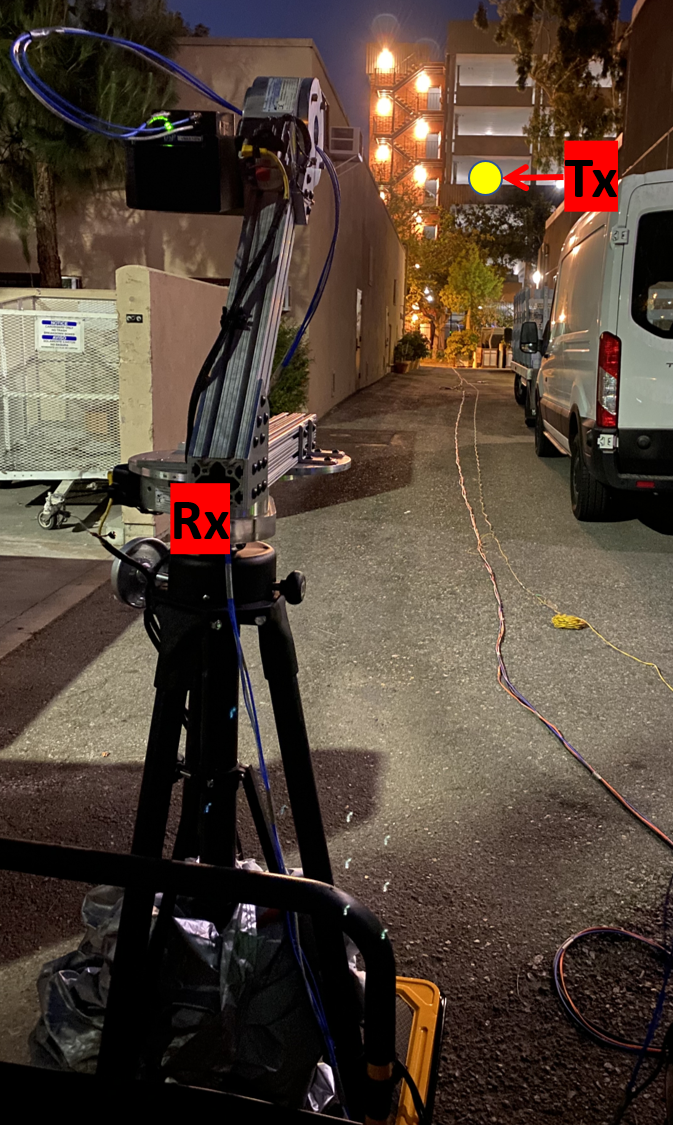}
		\caption{Tx4-Rx19 LoS $d=64.6 m$.}
		\vspace*{0mm}
		\label{fig:LOS TX4-RX19}
	\end{subfigure}
	\begin{subfigure}{0.3\textwidth}
		\centering
		\centering
		\hspace{0mm}
		\includegraphics[width=\columnwidth]{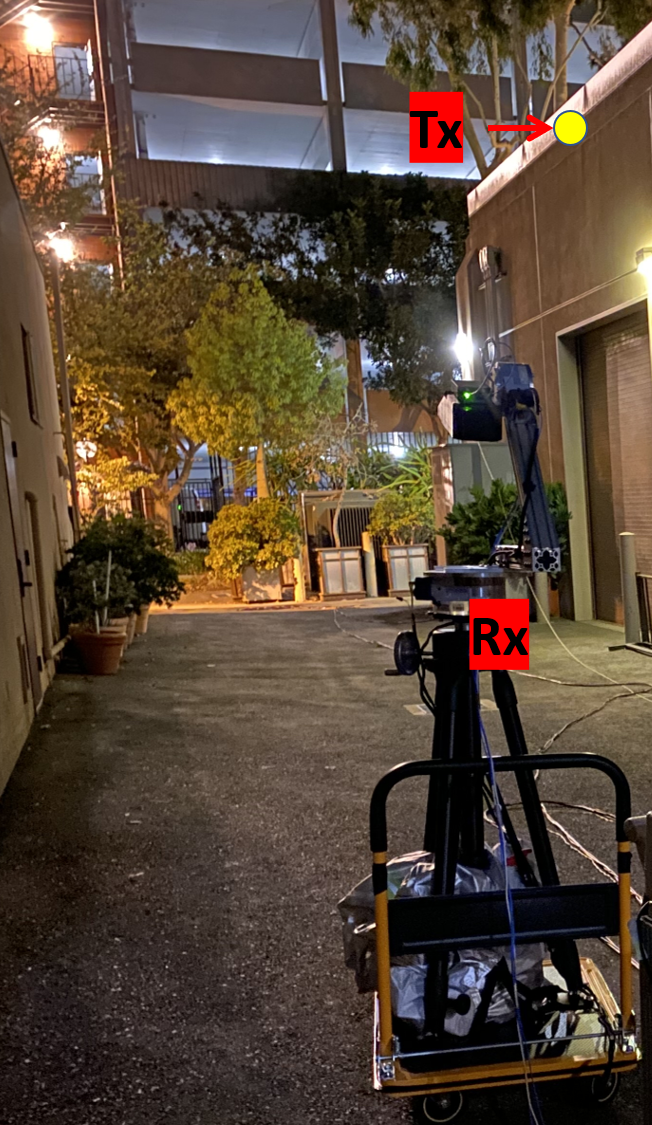}
		\caption{Tx5-Rx23 NLoS $d=45.5 m$.}
		\vspace*{0mm}
		\label{fig:NLOS TX5-RX23}
	\end{subfigure}
	\begin{subfigure}{0.3\textwidth}
		\centering
		\centering
		\hspace{0mm}
		\includegraphics[width=0.92\columnwidth]{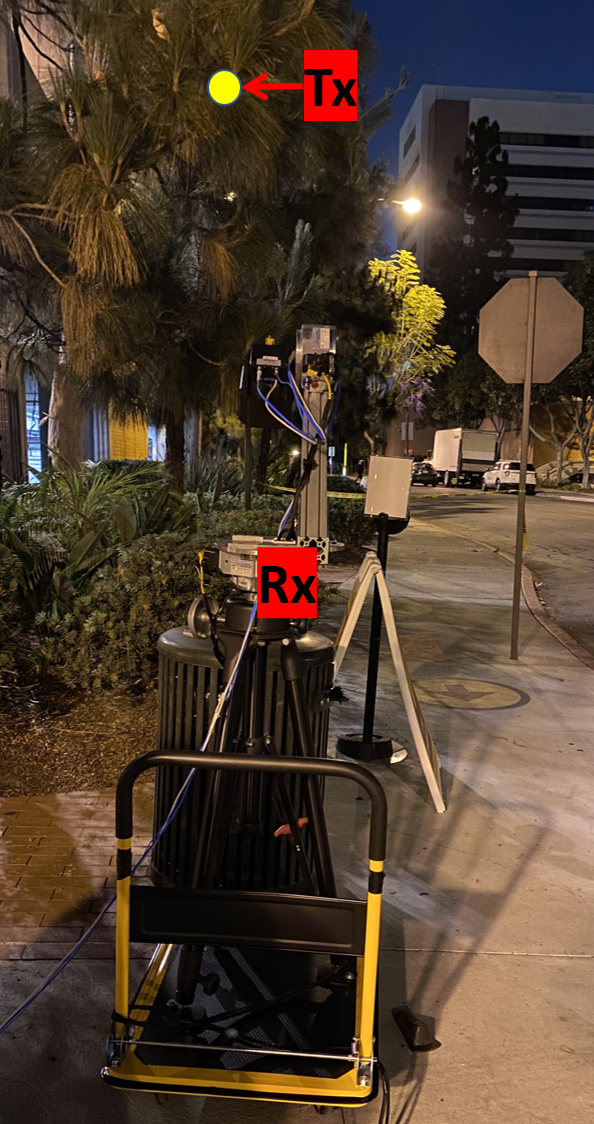}
		\caption{Tx6-Rx24 NLoS $d=20.4 m$}
		\label{fig:NLOS TX6-RX24}
	\end{subfigure}    
	\caption{LoS and NLoS sample points for Routes Four, Five and Six.}
	\label{fig:TX4_5}
\end{figure*}

Finally, for Route Six, the points are located on the McClintock side of PSA, approximately 10 meters behind the location of the Tx on Route One. The Rx locations were placed on the sidewalk next to McClintock Ave. Similar to the points in Route One, Olin Hall of Engineering (OHE), and RTH building create a "street canyon" environment for this route. The main obstruction of the LoS is provided by the foliage between the Tx and Rx locations. A sample point (Tx6-Rx24) is shown in Fig. \ref{fig:NLOS TX6-RX24}. Table \ref{tab:dist_Tx-Rx} shows a summary of the routes, locations and distances for all the measurement points of the campaign. 

\begin{table}[ht]
	\centering
	\caption{Description of Tx-Rx links and their respective direct distances.}
	\label{tab:dist_Tx-Rx}
	\begin{tabular}{|c|c|c|c|c|}
		\hline 
		\textbf{Tx identifier}          & \textbf{LoS Rx identifier}         & $\mathbf{d_{LoS}}$ \textbf{(m)}    & \textbf{NLoS Rx identifier}  & $\mathbf{d_{NLoS}}$ \textbf{(m)} \\ \hline \hline
		\multicolumn{1}{|l|}{\textbf{$Tx_1$}}           & 1-6               & 82.5, 64.5, 40.8, 72.3, 49.8, 32.1             & 7-9       & 83.2, 73.6, 46.4        \\ \hline
		\multicolumn{1}{|l|}{\textbf{$Tx_2$}}           & 10-13 & 20.4, 33.9, 45.9, 54.3 & -    & -               \\ \hline
		\multicolumn{1}{|l|}{\textbf{$Tx_3$}}           & - & -     & 14-16       & 62.6, 53.4, 40.7            \\ \hline
		\multicolumn{1}{|l|}{\textbf{$Tx_4$}}           &  17-19   &    36.3, 57.9, 65.7 & - & - 
		\\ \hline
		\multicolumn{1}{|l|}{\textbf{$Tx_5$}}           & -  & -  &  20-23  &   35, 58.5, 66.8, 45.5  \\ \hline
		\multicolumn{1}{|l|}{\textbf{$Tx_6$}}           & - & - & 24-26 & 20.8, 30,20           \\ \hline
	\end{tabular}
\end{table}

\section{Parameters and processing}

\subsection{Data processing}
The VNA-based measurement setup explained in Section II produces a collection of frequency scans for each Tx-Rx geographical location. Each measurement can be described as a five-dimensional tensor  $H_{meas}(f,\phi_{Tx},\tilde{\theta}_{Tx},\phi_{Rx},\tilde{\theta}_{Rx};d)$ where $f$ denotes the frequency points over the 1 GHz bandwidth (145-146 GHz), $\phi_{Tx}$ and $\phi_{Rx}$ denote the azimuth orientation of the Tx and Rx, respectively, $\tilde{\theta}_{Tx}$ and $\tilde{\theta}_{Rx}$ denote elevation orientation of the Tx and Rx, respectively, and $d$ is the Tx-Rx distance. Each tensor, $H_{meas}$, has dimensions of $N \times N^{\tilde{\theta}}_{Tx} \times N^{\phi}_{Tx} \times N^{\tilde{\theta}}_{Rx} \times N^{\phi}_{Rx}$ where $N$ is the number of frequency points per sweep (1001), $ N^{\tilde{\theta}}_{Tx}$ and $ N^{\tilde{\theta}}_{Rx}$ are the number of azimuth directions at the Tx (13) and Rx (36), and $ N^{\phi}_{Tx}$ and $ N^{\phi}_{Rx}$ are the number of elevation directions at the Tx $(3)$ and Rx $(3)$, respectively. Before the processing and parameter analysis we calibrate the measurement (eliminating the effects of the system and antennas) transfer functions. The OTA calibration $H_{OTA}(f)$ is used to obtain the calibrated directional channel transfer function by dividing the measured channel transfer function by the OTA calibration: $H(f,\phi_{Tx},\tilde{\theta}_{Tx},\phi_{Rx},\tilde{\theta}_{Rx};d) =H_{meas}(f,\phi_{Tx},\tilde{\theta}_{Tx},\phi_{Rx},\tilde{\theta}_{Rx};d)/H_{OTA}(f)$. 
The calibrated channel frequency response is used to compute different parameters such as the directional power delay profile (PDP) as
\begin{equation}
	P_{calc}(\tau,\phi_{Tx},\tilde{\theta}_{Tx},\phi_{Rx},\tilde{\theta}_{Rx},d)=|\mathcal{F}_{f}^{-1}\{H(f,\phi_{Tx},\tilde{\theta}_{Tx},\phi_{Rx},\tilde{\theta}_{Rx},d)\}|^2,
\end{equation}
where $\mathcal{F}_{f}^{-1}$ is the inverse fast Fourier transform (IFFT) with respect to $f$.  
To minimize the effects of noise, thresholding and delay gating are applied similar to \cite{gomez-ponce2020,abbasi2021double} that is expressed as 
\begin{equation}
	P(\tau)=[P_{calc}(\tau): (\tau\leq\tau_{gate})  \land (P_{calc}(\tau)\geq P_{\lambda})]
\end{equation}  
or $0$ if it does not fulfill these conditions. The value $\tau_{gate}$ is the delay gating threshold set to avoid using long delay bins or points with the "wrap-around" effect of the IFFT. $P_{\lambda}$ is the noise threshold that is selected to ignore the power of delay bins with noise which could particularly distort delay spread and angular spread. For the current measurements, $\tau_{gate}$ is set to 933.33 ns (corresponding to 280 m excess runlength) and $P_{\lambda}$ is selected to be 6 dB above the noise floor (average noise power) of the PDP.

From the collection of directional PDPs we selected the strongest beam as the beam-pair with the highest power (max-dir) as
\begin{equation}
	P_{\rm max}(\tau)=P(\tau,\phi_{\hat{i}},\tilde{\theta}_{\hat{j}},\phi_{\hat{k}},\tilde{\theta}_{\hat{l}},d); (\hat{i},\hat{j},\hat{k},\hat{l}) = \max_{i,j,k,l} \sum_\tau P(\tau,\phi_i,\tilde{\theta}_j,\phi_k,\tilde{\theta}_l,d).
\end{equation}
Finally, an "omni-directional" PDP is constructed by first combining all the elevations by summing over different elevations for each delay bin, and then selecting the azimuth with the strongest contribution. The selection of the strongest azimuth direction per delay bin to reconstruct a PDP is similar to \cite{Hur_omni,abbasi2021ultra}. Overall, this process can be summed up as 

\begin{equation}
	P_{\rm omni}(\tau;d)=\max_{\phi_{Tx},\phi_{Rx}} \sum_i \sum_j P(\phi_{Tx},\tilde{\theta}_{Tx}^i,\phi_{Rx},\tilde{\theta}_{Rx}^j;d).
\end{equation}
where $i,j\in  \{1,2,3\}$ represents the elevations ($ \tilde{\theta}_{Tx}^i, \tilde{\theta}_{Rx}^j \in \{-13^\circ,0^\circ,13^\circ\} $) for Tx and Rx, respectively. The adding of the different elevation cuts is meaningful because the spacing of the cuts in the elevation domain was taken as $13^\circ$, which is identical to the (full width half maximum (FWHM)) beamwidth. Thus, the effective elevation pattern of the sum is approximately constant in the range $-13^\circ \le \tilde{\theta}_{Tx} \le 13^\circ$, and has a FWHM of $39^\circ$, and similar at the Rx
%It is important to mention that the previous formula only considers a {\em single} elevation. To combine all of the elevations, we sum over the different elevations for each delay bin, and then selected the azimuth with the strongest contribution. This is valid since each elevation capture is done at angles equal to $13^\circ$ (HPBW) 

%\begin{equation}
	%\label{eq:omni}
	%P_{\rm omni}(\tau;d)=  \sum_i \sum_jP_{\rm omni}^{i,j}(\tau;d).
	%\end{equation}

\subsection{Parameter computation}
\label{sec:par}
Similar to the analysis performed in \cite{Abbasi2021THz}, we use the directional and omni-directional PDPs described in the previous section to compute several condensed parameters in order to characterize the propagation channels. The computations are based on the noise-thresholded and delay-gated PDPs calculated as described above. 
\subsubsection{Path loss and shadowing}
The first parameter to be computed is the path loss. By definition (\cite{molisch2012wireless}) it is computed as the sum of the power on each delay bin in the PDP.
\begin{equation}
	PL_i(d)=\sum_\tau P_i(\tau,d),
\end{equation}
where $i$ can denote omni-directional (omni) or the strongest beam (best-dir).
To model its behavior as a function of distance, we use the classical single slope "power law" also known as $\alpha - \beta$ model, such that the pathloss in dB is 
\begin{equation}
	PL_{\rm dB}(d)=\alpha+10\beta \log_{10}(d)+\epsilon,
\end{equation}
where $\alpha$ and $\beta$ are the estimated parameters, and $\epsilon$ represents the "Shadowing" or random variation of the data with respect to its mean. It is assumed to follow a zero-mean normal distribution $\epsilon \sim N(0,\sigma)$, where $\sigma$ is the standard deviation of the distribution. To obtain the parameters of the model, we can use approaches such as maximum likelihood estimation (MLE) or ordinary least squares (OLS) \cite{molisch2012wireless,kartunen_PL}. Following common assumptions in the modeling of path loss, the procedure is separated between the ensemble of LoS and NLoS measurement points.

An analysis carried out in \cite{karttunen2016path} describes the challenges of an uneven density of distances between the Tx and Rx (in linear and logarithmic scale). This non-uniformity can lead to an increasing in the leverage of some points in the regression analysis compared to others. To compensate for this effect, \cite{karttunen2016path} implemented a weighted regression model for path loss modeling. Each weight ($w_i$) is computed according to the density of points along  the distance in $log_{10}$ scale. So, $w_i$ will be larger for points located in low density areas and vice versa. While multiple weighting methods are described in the paper, however, we adopt the approach of "equal weights to N bins over $log_{10}(d)$ ($w_i \propto log_{10}(d)$)", because this strategy corresponds to a least square fitting of "dB vs $log_{10}(d)$".
\subsubsection{Delay spread}
The rms delay spread (RMSDS) is calculated as the second central moment of the PDP \cite{molisch2012wireless}:
\begin{equation}
	\sigma_\tau=\sqrt{\frac{\int_\tau P_i(\tau)\tau^2 d\tau}{\int_\tau P_i(\tau)d\tau} - \left(\frac{\int_\tau P_i(\tau)\tau d\tau}{\int_\tau P_i(\tau)d\tau}\right)^2},
\end{equation}
where $i$ can be "omni" or "max-dir". Noise and delay thresholding are essential for reducing the impact of long-delayed artefacts. Since this parameter is defined for continuous waveforms, therefore to approximate it, we increase the number of samples in the PDPs by oversampling them. Additionally, we apply a Hann window to reduce the impact of the sidelobes in the parameter estimation.  
\subsubsection{Angular spread}
\label{sect:AS}
The measurement campaign creates a "virtual" MIMO scenario for each location pair,  allowing angular analysis. A way to quantify the dispersion of power over different angular directions is the angular spread. The starting point of its computation is the double-directional angular power spectrum ($DDAPS_{full}$), a function of the power concentration over different directions (particular azimuth, elevation directions) at Tx and Rx.
The DDAPS is computed as 
\begin{equation}
	DDAPS_{full}(\phi_{Tx},\tilde{\theta}_{Tx},\phi_{Rx},\tilde{\theta}_{Rx};d)=\sum_\tau P(\tau,\phi_{Tx},\tilde{\theta}_{Tx},\phi_{Rx},\tilde{\theta}_{Rx};d).
\end{equation}
Similar to the delay spread analysis, noise and delay gating are important before the computation of $DDAPS_{full}$ to minimize noise accumulation in directions where no significant MPC is observed. Using the $DDAPS_{full}$, we add the contribution of different elevations from both ends to have a similar DDAPS as \cite{Abbasi2021THz}.
\begin{equation}
	DDAPS(\phi_{Tx},\phi_{Rx};d)=\sum_{\tilde{\theta}_{Tx}}\sum_{\tilde{\theta}_{Rx}} DDAPS_{full}(\phi_{Tx},\tilde{\theta}_{Tx},\phi_{Rx},\tilde{\theta}_{Rx};d).
\end{equation}
We combine the different elevations we measured since the limited number of elevation cuts (which was imposed by limits on the measurement duration) is insufficient for a detailed elevation analysis. Moreover, since the direction of the primary propagation is well covered, it is expected that there will be less information in other elevation cuts. 
%We combine the different elevations because of the insufficient number of elevation points to perform an elevation analysis, additionally each elevation scan was taken in steps of $13^\circ$ (HPBW). 
Finally, to compute the (azimuthal) angular power spectrum (APS) at the Tx, we integrate over $\phi_{Rx}$, and do the same for the APS at the Rx. Using the APS, we compute the angular spread by applying Fleury's definition \cite{fleury2000first}:
\begin{equation}
	\sigma^\circ=\sqrt{\frac{\sum_\phi \left|e^{j\phi}-\mu_\phi \right|^2 APS_k(\phi)}{\sum_\phi APS_k(\phi)}},
\end{equation}
where $k$ can be Tx or Rx indicating departure or arrival APS and $\mu_\phi$ can be computed as
\begin{equation}
	\mu_\phi=\frac{\sum_\phi e^{j\phi} APS_k(\phi)}{\sum_\phi APS_k(\phi)}.
\end{equation}
It is important to mention that the obtained values will be an upper bound for the actual angular spreads of the channel due to the finite horn antenna beamwidth \cite{Abbasi2021THz}.

\subsubsection {Power distribution over MPC}
In channel analysis, it is important to examine the power distribution of MPCs over the delay domain. Specially, the concentration of power in the "strongest" MPC versus the rest of the MPCs in the channel. Thus, we define $\kappa_1$, a parameter computed as follows: 
\begin{equation}
	\kappa_1=\frac{P_i(\tilde{\tau}_1)}{\sum_{\tilde{\tau}=\tilde{\tau}_2}^{\tilde{\tau}_N} P_i(\tilde{\tau})},
\end{equation}
where $i$ can be "omni" or "max-dir", and $\tilde{\tau_k}$ is the delay bin of the $k$-th local maximum of the PDP $P_i(\tilde{\tau})$, ordered by magnitude, so that $\tilde{\tau_1}$ signifies the location of the largest local maximum.

As explained in \cite{abbasi2020channel}, $\kappa_1$ is different from the "Rice Factor" because it is not possible to differentiate between closely spaced MPCs, therefore, the local maximum of the PDP is not strictly identical to an MPC. To perform the most accurate Rice Factor analysis, HRPE can be used so that MPCs are properly identified, and this will be presented in future work. Similarly as $\sigma_\tau$ we apply oversampling and a Hann window to avoid the sidelobe effects and to have a better estimation of the parameter.\\
In the next section, regression analysis will be added in the estimation of the parameters $\sigma_\tau,\kappa_1$ similar to \cite{Abbasi2021THz}. With this regression, we will observe their behavior with respect to the distance between Tx and Rx. The linear regression model is with respect to logarithmic quantities and it is $Z=\alpha+\beta \log_{10}(d)$.

\section{Measurement results}
In this section the results for the measurement campaign are discussed. 

\subsection{Power delay profiles}
\begin{figure}
	\centering
	\centering
	\hspace{7mm}
	\includegraphics[width=0.5\columnwidth]{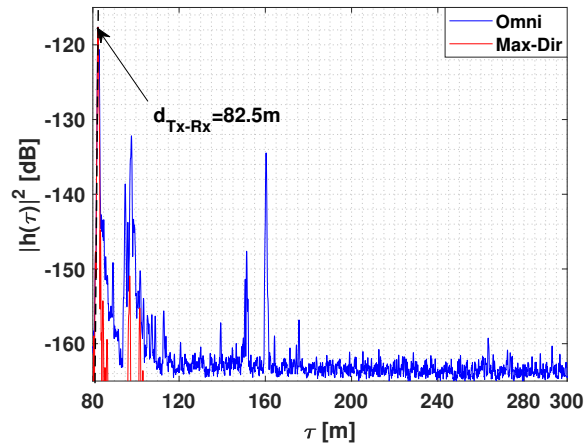}
	\caption{LoS case with $d=82.5 m$ (Tx1-Rx1).}
	\label{fig:PDP-LOS}
\end{figure}

To start with the measurement analysis, we first present some sample PDPs, characterizing one LoS and two NLoS location pairs. The LoS measurement was taken at a distance of 82.5 m. Fig. \ref{fig:PDP-LOS} presents the omni-directional and max-dir PDPs. The LoS MPC is clearly observed in both the max-dir and omni-directional PDPs. Apart from the LoS MPC, multiple MPCs with runlength $\leq 160$ m with power only up to 30dB lower than the LoS. These "extra" components are diminished in the max-dir as a result of the spatial filtering effect provided by the antennas. In this particular case, for the omni-directional case, we observed several (very weak) MPCs arriving before the LoS MPC. As explained in section II, the maximum measurable excess delay of the system is $1 \mu s$ which leads to 300 m of maximum runlength. Any MPC with delay $\geq 1 \mu s$ will suffer from aliasing, and so be wrapped around in delay domain. This effect was corrected for all figures. Additionally, the PDPs shown are oversampled and windowed using a Hann window to diminish the effect of sidelobes and observe low power MPCs. \\    

For the NLoS case, we present two location pairs, with Tx-Rx distances of 45.5 and 83 m, respectively. A richer multipath scenario is expected because of the attenuation of the LoS component and increase of additional MPCs that arrive at the Rx. In the case of the 45.5 m measurement, we see a concentrated max-dir PDP, and small quantity of additional MPCs with power $\leq 30dB$, similar to a LoS scenario. The scenario for this measurement is shown in Fig. \ref{fig:NLOS TX5-RX23}, and as can be seen, the Tx is set in the PSA building and the Rx is located in the alley between TTL and SCD, creating a "street-canyon" and concentrating (in the delay domain) the power reaching to the Rx, since all components guided by the canyon have fairly similar delays created by different number of reflections on the housewalls, which are just a street width apart. We also note that while the first pronounced peak in the PDP is the strongest one, it is {\em not} a quasi-LoS (as often observed at low frequencies), as shown by the fact that its associated delay is {\em longer} than that of the (theoretical) LoS.

\begin{figure*}[t!]
	\centering
	\begin{subfigure}{0.45\textwidth}
		\centering
		\hspace{7mm}
		\includegraphics[width=1\columnwidth]{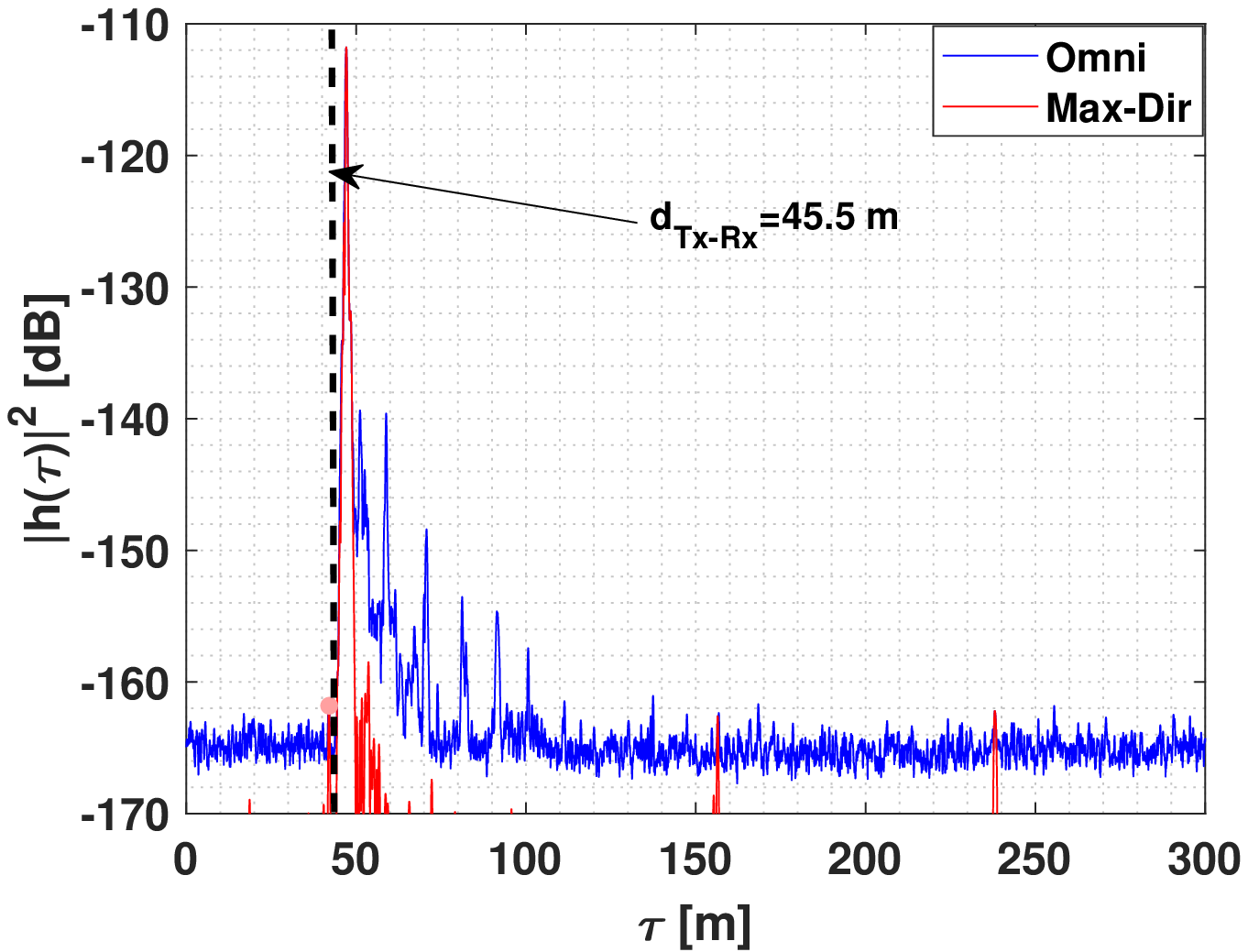}
		\caption{NLoS case with $d=45.5 m$ (Tx5-Rx23).}
		\label{fig:PDP-NLOS1}
	\end{subfigure}
	\begin{subfigure}{0.45\textwidth}
		\centering
		\centering
		\hspace{7mm}
		\includegraphics[width=1\columnwidth]{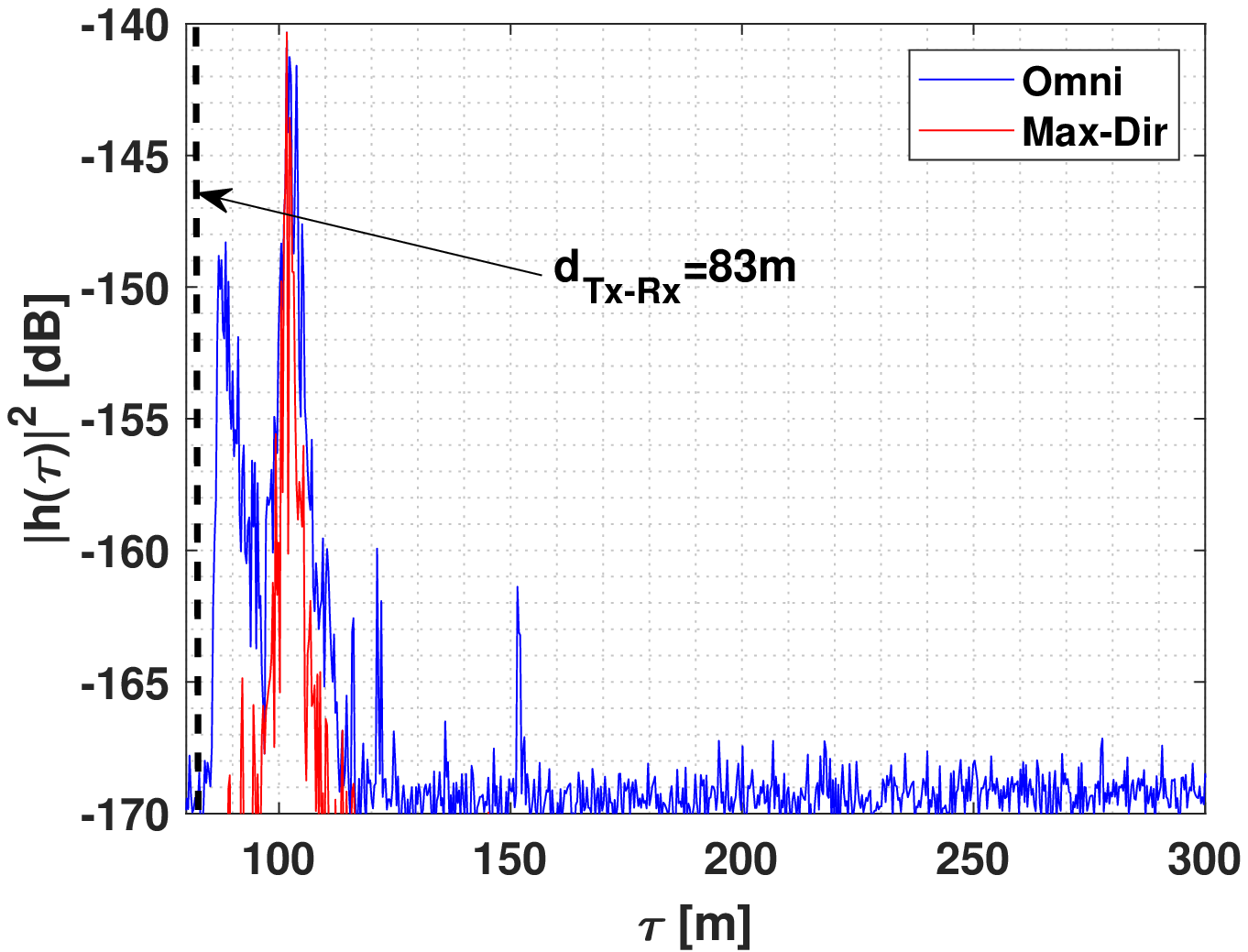}
		\caption{NLoS case with $d=83 m$ (Tx1-Rx7).}
		\vspace*{0mm}
		\label{fig:PDP-NLOS2}
	\end{subfigure}
	\caption{PDP for two sample NLoS measurement cases.}
	\label{fig:PDP-NLOS}
\end{figure*}

The second point is shown at Fig. \ref{fig:NLOS TX1-RX7}, in this case it is observed that there is a larger set of MPCs, especially for the omni-directional PDP, compared to the previous NLoS case. These MPCs are a product of reflections coming from the RTH building. This effect can be noticed in Fig. \ref{fig:APS-NLOS2}, and we see that the first significant MPC is not the strongest one. More details of this scenario will be discussed in the next subsection in more detail.

\subsection{Angular power spectrum}

This section discusses the Angular Power Spectrum (APS) of the selected sample LoS and NLoS location pairs. For the LoS case, we observe a large concentration of MPCs in the LoS direction, an additional concentration of MPCs can be observed at $\phi_{Tx}=37,\phi_{Rx}=35$. These MPCs correspond to reflections coming off the RTH building, additionally, we can also observe MPCs at angles close to $\phi_{Tx}=0,\phi_{Rx}=180$.\\

\begin{figure}
	\centering
	\centering
	%\hspace{7mm}
	\includegraphics[width=0.5\columnwidth]{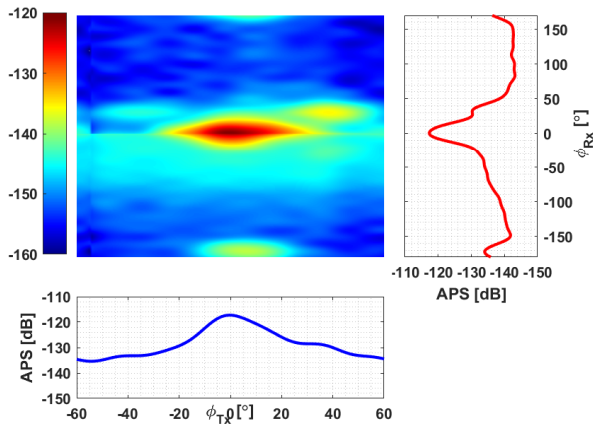}
	\caption{LoS APS for $d=82.5 m$ (Tx1-Rx1).}
	\vspace*{0mm}
	\label{fig:APS-LOS}
\end{figure}

\begin{figure*}
	\centering
	\begin{subfigure}{0.45\textwidth}
		\centering
		%        \hspace{7mm}
		\includegraphics[width=1\columnwidth]{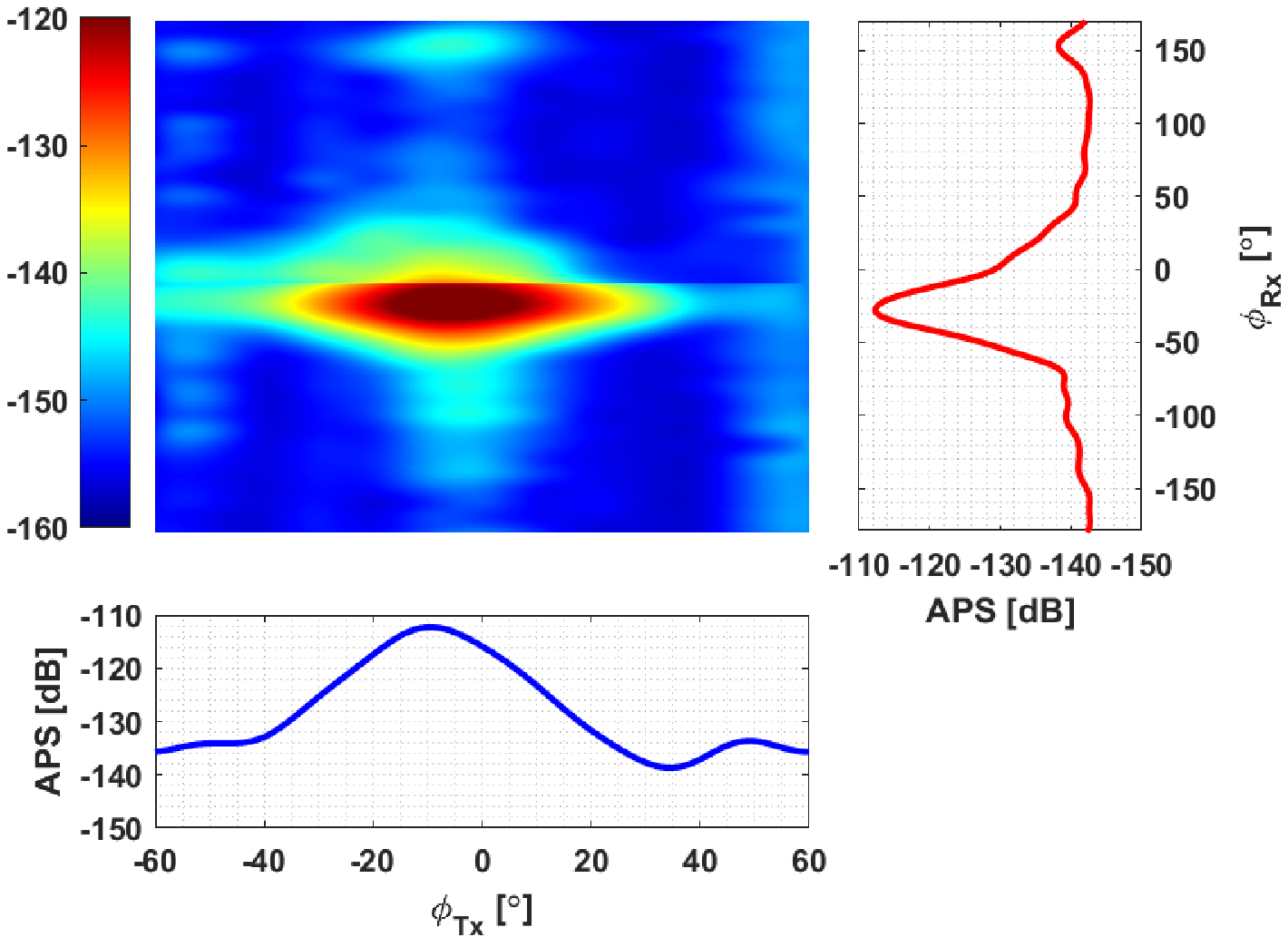}
		\caption{NLoS APS for $d=45.5 m$ (Tx5-Rx23).}
		\vspace*{0mm}
		\label{fig:APS-NLOS1}
	\end{subfigure}
	\begin{subfigure}{0.45\textwidth}
		\centering
		%        \hspace{7mm}
		\includegraphics[width=1\columnwidth]{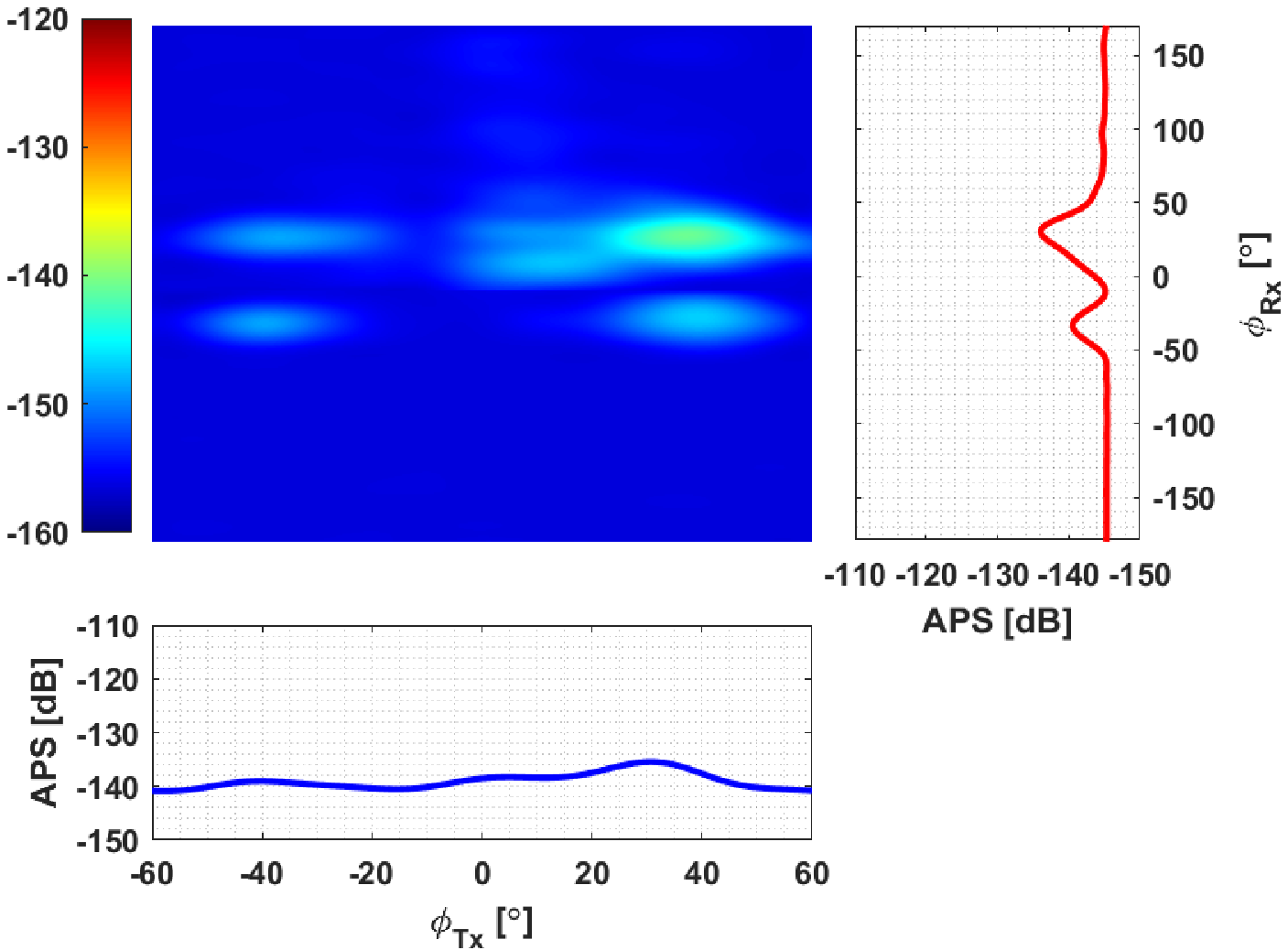}
		\caption{NLoS APS for $d=83 m$ (Tx1-Rx7).}
		\vspace*{0mm}
		\label{fig:APS-NLOS2}
	\end{subfigure}
	\caption{Sample NLoS APSes for two cases.}
	\label{fig:APS-NLOS}
\end{figure*}

The NLoS points have a different behavior compared to LoS. In the case of the point Tx5-Rx20, we see a large concentration of MPCs in one main direction, similar as in the sample LoS. However, the center of this concentration is not not in the LOS direction, but rather in the direction of the street, with  $\phi_{Tx}=-15,\phi_{Rx}=-27$. 
%, with similar power concentration as seen in the sample LoS.
%, but less than half of the distance of the sample LoS point (Tx1-Rx1, $d=82.5$ m). 
This concentration of MPCs are a product of the "street canyon" effect created by the SCD and the TTL building (see Fig. \ref{fig:NLOS TX5-RX23}). An additional concentration of MPCs can be observed at $\phi_{Tx}=-15,\phi_{Rx}=47$; in this case, the Tx horn is still facing towards the canyon but the receiver collects a weaker reflection inside it. For the last NLoS location pair, (Tx1-Rx7) shows,
has a distance of ($d=83$ m), and 
as can be seen in Fig. \ref{fig:APS-NLOS2}, several maximuma in the APS, with the strongest one at $\phi_{Tx}=37,\phi_{Rx}=28$. This corresponds to Tx and Rx looking towards the RTH building, and is thus congruent with the scenario observed in Fig. \ref{fig:NLOS TX1-RX7}. As can be seen in the picture, the LoS is blocked by the pillars in front of the receiver and the right-hand side of the receiver has an opening facing towards McClintock Ave, the OHE, RTH and EEB buildings. Moreover, additional weaker MPCs (approx. 8dB weaker than the strongest MPCs) are observed at $\phi_{Tx}=-38,\phi_{Rx}=27$. These MPCs are reflections from RTH, similar to the previous MPCs, however they reach the receiver from the left hand side gap observed between the inner wall of GER building and the pillar, which means additional attenuation.  

The above discussions not only provide a description of relevant propagation effects, but also support the correctness of the measurements, as the extracted MPCs are in agreement with the geometry of the environment. Further verifications, not shown here for space reasons, were done for other location pairs as well.

\subsection{Path loss and shadowing}

\begin{figure*}[t!]
	\begin{subfigure}[b]{0.45\textwidth}
		\centering
		%\hspace{7mm}
		\includegraphics[width=1\columnwidth]{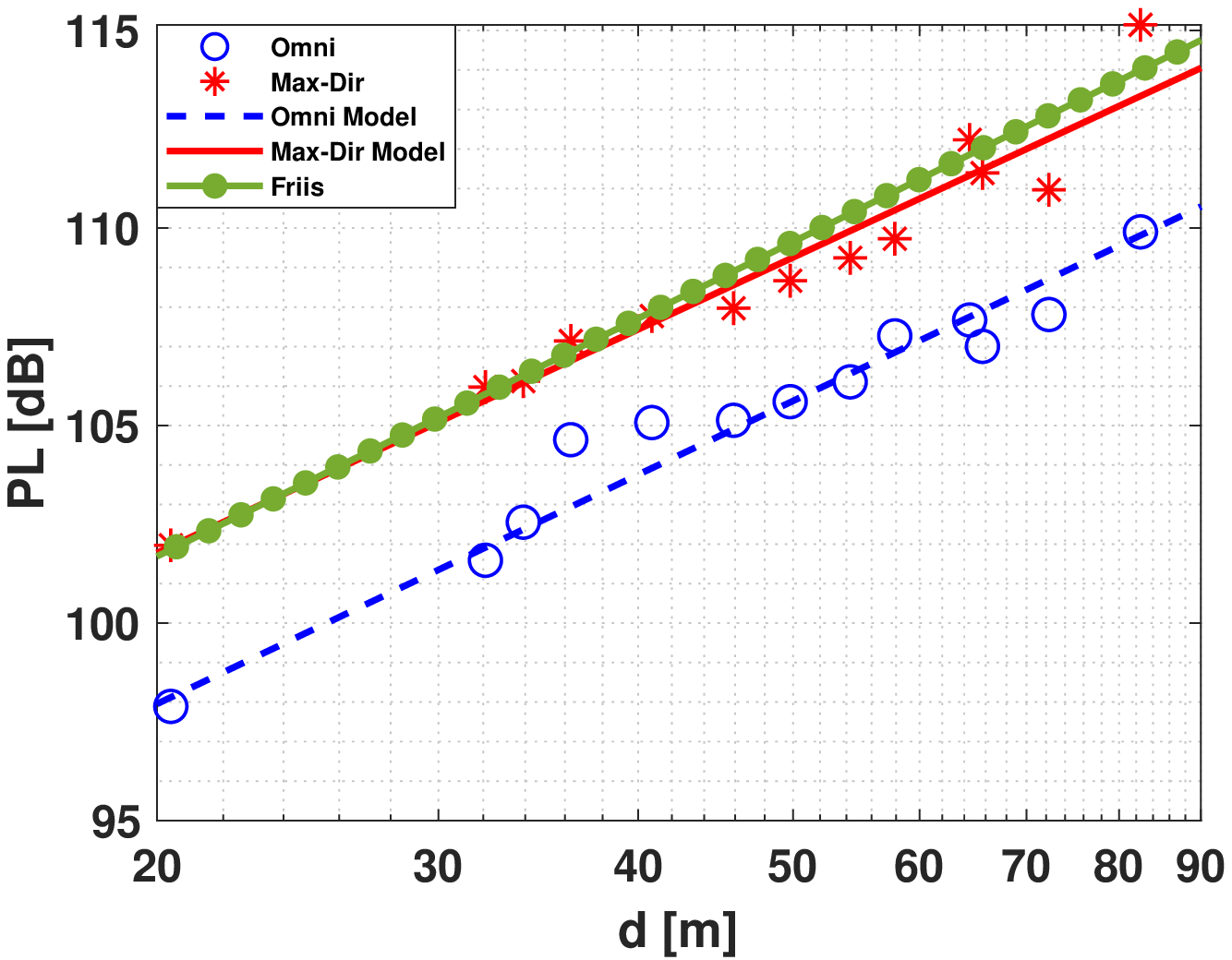}
		\caption{Path loss modeling with $log_{10}(d)$ weighting.}
		\vspace*{0mm}
		\label{PLOSS-LOS}
	\end{subfigure}
	\centering
	\begin{subfigure}[b]{0.45\textwidth}
		\centering
		%\hspace{7mm}
		\includegraphics[width=1\columnwidth]{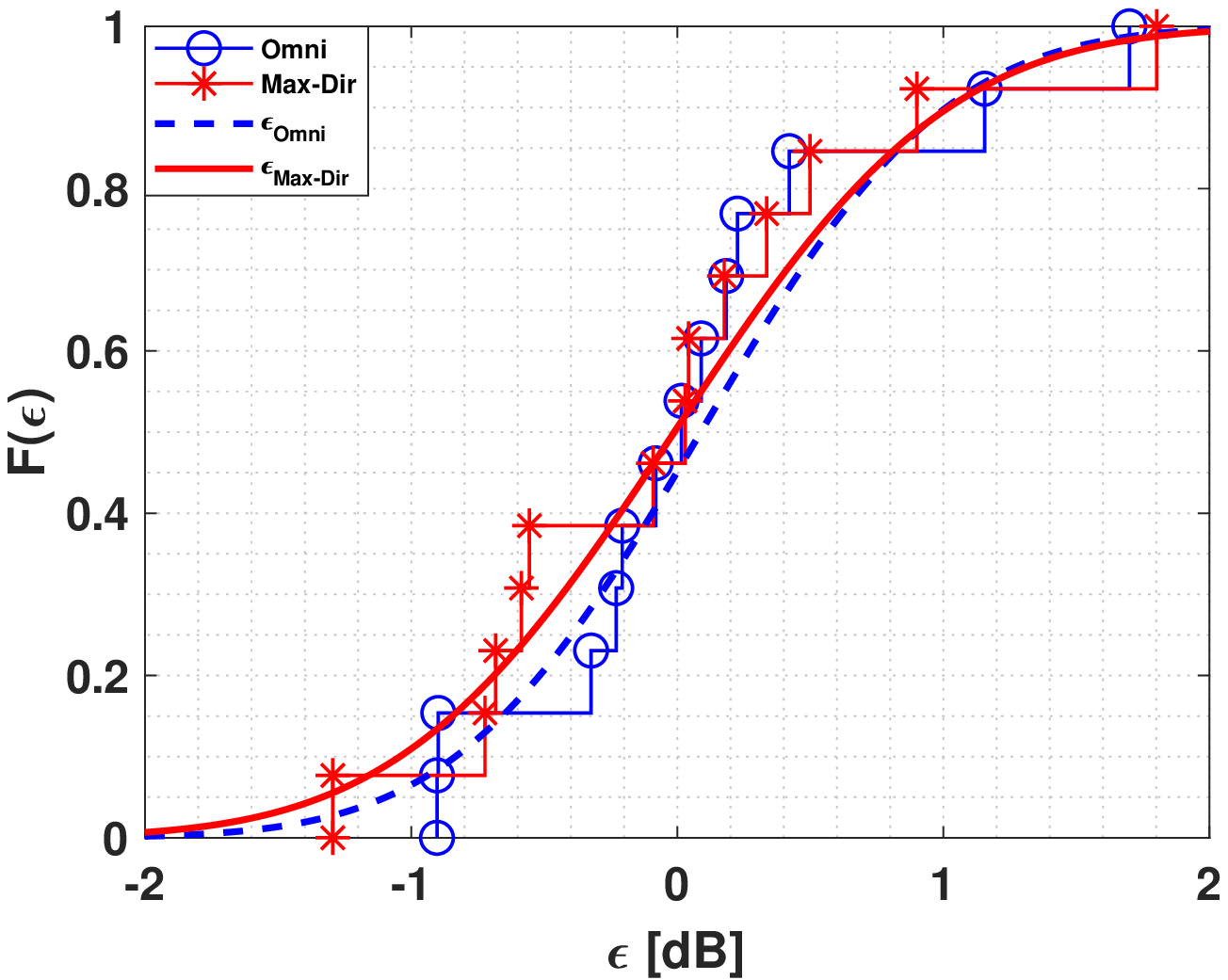}
		\caption{Shadowing.}
		\vspace*{0mm}
		\label{SHA-LOS}
	\end{subfigure}
	\caption{Path loss and shadowing models for LoS points.}%
	\label{fig:los_PL_SHA}%
	\vspace{-0 mm}
	\vspace{-5mm}
\end{figure*}

In this section we start analyzing the ensemble of measurement locations. For the analysis, the points will be separated into LoS and NLoS to analyze their characteristics separately. For the LoS case, Fig. \ref{PLOSS-LOS} shows the path loss analysis using "max-dir", "omni-directional" PDPs and the Friis Model. For all points it can be observed that the path loss for the "max-dir" is larger or equal to the "omni" path loss points ($PL_{max-dir}\geq PL_{omni}$). Max-dir and omni-directional PL models are lower than the Friis model. The PL exponent is $\beta=1.88$, lower than the free space model. This fact is congruent with the scenario because the LoS points in Routes One and Four are in "street canyon" LoS environments (9 of 13 locations), therefore the "waveguiding" effect will produce a path loss lower than the free space. The parameters extracted by the "weighted" regression and the OLS are similar because of the low variations of the points against their linear models, additionally the shadowing shows the same variance in both cases and has a small difference in the mean value. 

\begin{figure*}[t!]
	\centering
	\begin{subfigure}[b]{0.45\textwidth}
		\centering
		%        \hspace{7mm}
		\includegraphics[width=1\columnwidth]{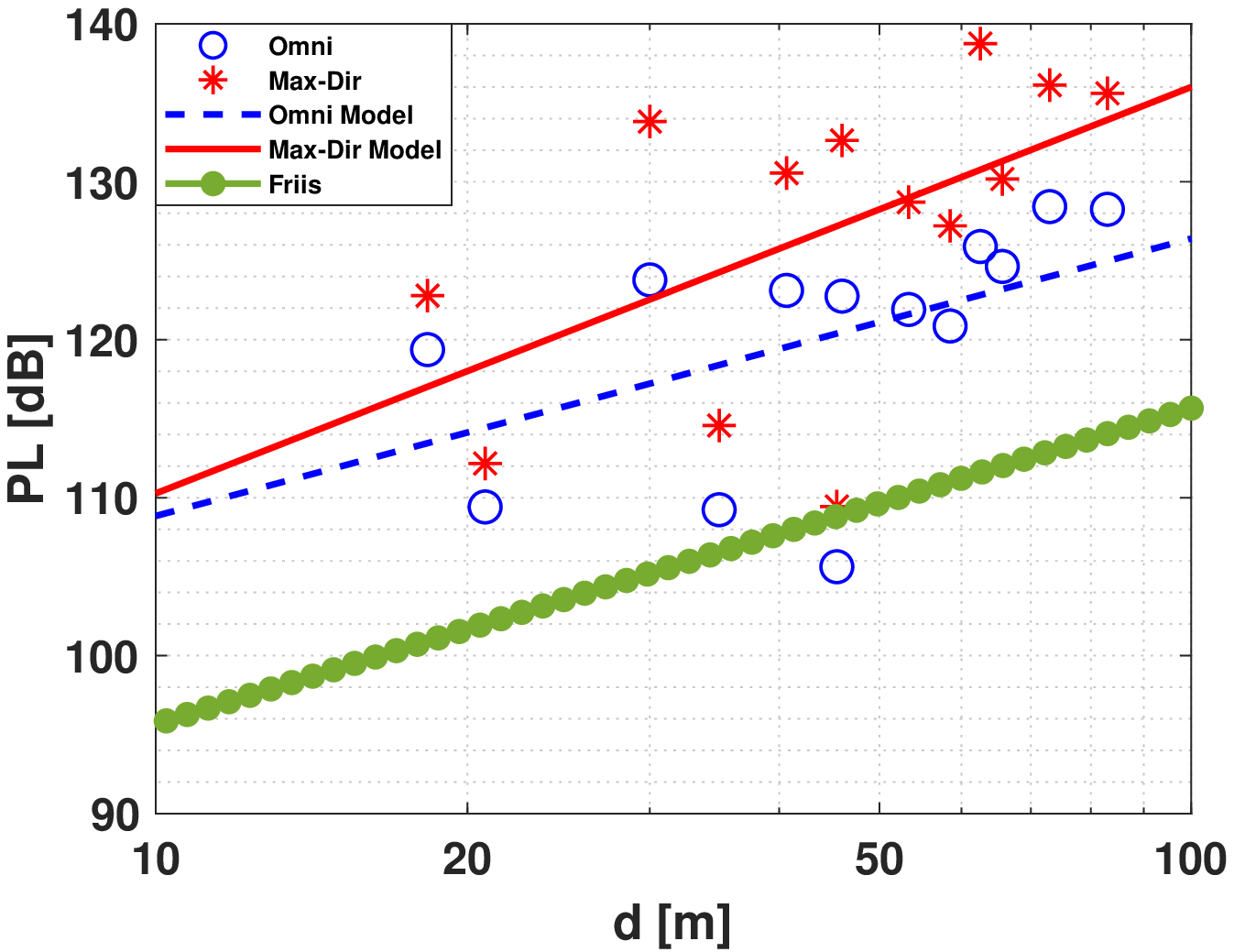}
		\caption{Linear fitting with $log_{10}(d)$ weighting.}
		\label{PLOSS-NLOS}
	\end{subfigure}
	\begin{subfigure}[b]{0.45\textwidth}
		\centering
		%        \hspace{7mm}
		\includegraphics[width=1\columnwidth]{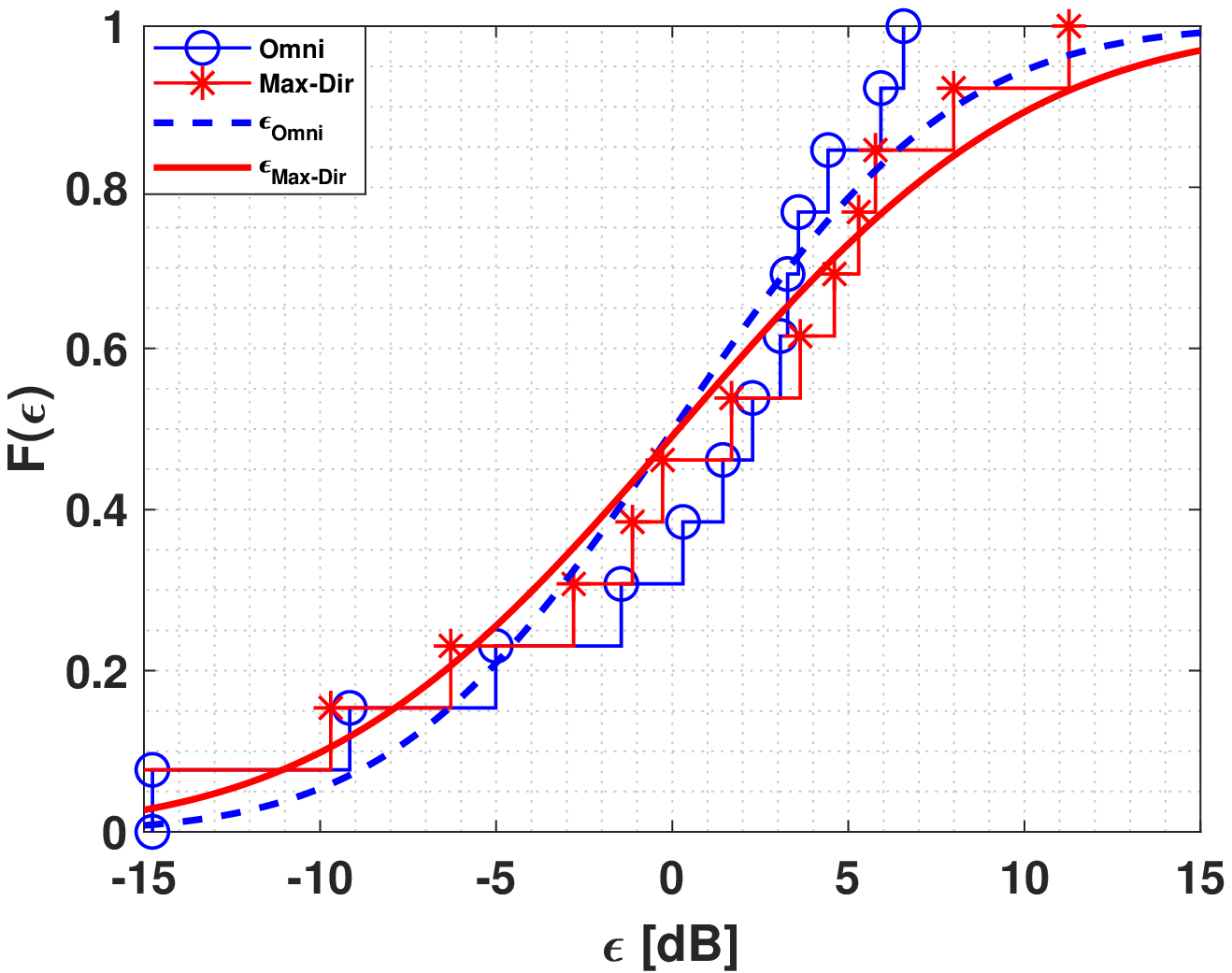}
		\caption{Shadowing.}
		\label{SHA-NLOS}
	\end{subfigure}
	\caption{Path loss and shadowing models for NLoS points.}%
	\vspace{-0 mm}
	\label{fig:nlos_PL_SHA}%
	\vspace{-5mm}
\end{figure*}

Fig. \ref{PLOSS-NLOS} shows the regression modeling for the NLoS case. The max-dir points show large values of PL compared to the omni-directional points, since in this case a significant percentage of energy is contained in MPCs whose directions are different from the max-Dir horn orientations. For a similar reason, the path loss exponent for the max-dir and omni-directional case are different ($\beta=2.57,\beta=1.76$ respectively). The omni-directional case has a smaller slope due to more MPCs from different directions provide energy at large distances. The shadowing oscillates between -15 and 15 dB for the omni and max-dir cases. The observed shadowing  standard deviations for both cases are 6.21 and 7.89 for the max-dir and omni-directional cases, respectively. A summary of the estimated regression parameters for path loss and statistical parameters for the shadowing with their respecting 95\% confidence interval is shown in Tables \ref{tab:PL} and \ref{tab:sha}.

In the NLoS case, we observed path loss values larger compared to Friis, except for the point (Tx5-Rx23). This point is located in a corridor between SCD and TTL buildings,  (see Fig. \ref{fig:NLOS TX5-RX23}). 
%The waveguiding effect provides a concentration of MPCs increasing the received power at the other end.\\
In this case there exists a very strong reflection, and the associated directional pathloss equals Friis, while the omni-directional pathloss is lower due to the existence of additional MPCs; similar to the LoS situation; this is {\em not} unphysical.

\begin{table*}[t!]
	\centering
	\caption{Path loss parameters with $95\%$ confidence interval.}
	\label{tab:PL}
	{%
		\begin{tabular}{|c|c|c|c|c|c|c|c|}
			\hline
			\multicolumn{1}{|c|}{\multirow{2}{*}{\textbf{Parameter}}} & \multicolumn{6}{c|}{\textbf{Linear model parameters estimated with 95\% CI}} \\ \cline{2-7} 
			\multicolumn{1}{|l|}{} & \multicolumn{1}{c|}{$\alpha$} & \multicolumn{1}{c|}{$\alpha_{min,95\%}$} & \multicolumn{1}{c|}{$\alpha_{max,95\%}$} & \multicolumn{1}{c|}{$\beta$} & \multicolumn{1}{c|}{$\beta_{min,95\%}$} & \multicolumn{1}{c|}{$\beta_{max,95\%}$} \\ \hline \hline
			$PL_{omni}^{LoS}$ & 72.88 & 69.91 & 75.86 & 1.93 & 1.74 & 2.11 \\ \hline
			$PL_{max-dir}^{LoS}$ & 77.33 & 74.1 & 80.57 & 1.88 & 1.68 & 2.08 \\ \hline
			$PL_{omni}^{LoS} OLS$ &  75.02 & 70.47 & 79.58 & 1.8 & 1.53 & 2.08 \\ \hline
			$PL_{max-dir}^{LoS} OLS$ & 77.06 & 71.74 & 82.37 & 1.89 & 1.58 & 2.21 \\ \hline
			$PL_{omni}^{NLoS}$ & 91.28 & 62.71 & 119.85 & 1.76 & -0.05 & 3.56 \\ \hline
			$PL_{max-dir}^{NLoS}$ & 84.54 & 49.21 & 119.88 & 2.57 & 0.34 & 4.81 \\ \hline
			$PL_{omni}^{NLoS} OLS$ & 86.81 & 52.96 & 120.66 & 2.03 & -0.01 & 4.07 \\ \hline
			$PL_{max-dir}^{NLoS} OLS$ & 82.91 & 39.96 & 125.87 & 2.68 & 0.09 & 5.27 \\ \hline
		\end{tabular}%
	}
\end{table*}

\begin{table*}[t!]
	\centering
	\caption{Shadowing model parameters with $95\%$ confidence interval.}
	\label{tab:sha}
	{%
		\begin{tabular}{|c|c|c|c|c|c|c|}
			\hline
			\multicolumn{1}{|c|}{\multirow{2}{*}{\textbf{Parameter}}} & \multicolumn{6}{c|}{\textbf{Statistical model parameters estimated with 95\% CI}} \\ \cline{2-7} 
			\multicolumn{1}{|l|}{} & \multicolumn{1}{c|}{$\mu$} & \multicolumn{1}{c|}{$\mu_{min,95\%}$} & \multicolumn{1}{c|}{$\mu_{max,95\%}$} & \multicolumn{1}{c|}{$\sigma$} & \multicolumn{1}{c|}{$\sigma_{min,95\%}$} & \multicolumn{1}{c|}{$\sigma_{max,95\%}$} \\ \hline \hline
			$\epsilon_{omni}^{LoS}$ & 0.09 & -0.35 & 0.52 & 0.72 & 0.52 & 1.19 \\ \hline
			$\epsilon_{max-dir}^{LoS}$ & -0.01 & -0.5 & 0.48 & 0.8 & 0.58 & 1.33 \\ \hline
			$\epsilon_{omni}^{LoS} OLS$ & 0 & -0.42 & 0.42 & 0.69 & 0.49 & 1.14 \\ \hline
			$\epsilon_{max-dir}^{LoS} OLS$ & 0 & -0.49 & 0.49 & 0.8 & 0.58 & 1.33 \\ \hline
			$\epsilon_{omni}^{NLoS}$ & 0.04 & -3.73 & 3.81 & 6.24 & 4.48 & 10.3 \\ \hline
			$\epsilon_{max-dir}^{NLoS}$ & 0.18 & -4.59 & 4.94 & 7.89 & 5.66 & 13.02 \\ \hline
			$\epsilon_{omni}^{NLoS} OLS$ & 0 & -3.76 & 3.76 & 6.21 & 4.46 & 10.26 \\ \hline
			$\epsilon_{max-dir}^{NLoS} OLS$ & 0 & -4.77 & 4.77 & 7.89 & 5.65 & 13.02 \\ \hline
		\end{tabular}%
	}
\end{table*}

\subsection{RMSDS}

\begin{figure*}[t!]
	\centering
	\begin{subfigure}[b]{0.45\textwidth}
		\centering
		%\hspace{7mm}
		\includegraphics[width=1\columnwidth]{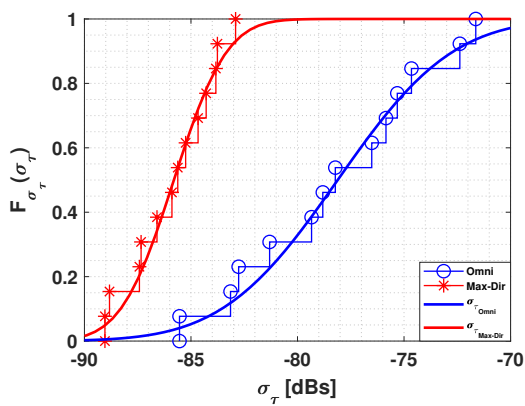}
		\caption{CDF of delay spread.}
		\vspace*{0mm}
		\label{fig:RMSDS-LOS-CDF}%
	\end{subfigure}
	\begin{subfigure}[b]{0.45\textwidth}
		\centering
		%\hspace{7mm}
		\includegraphics[width=1\columnwidth]{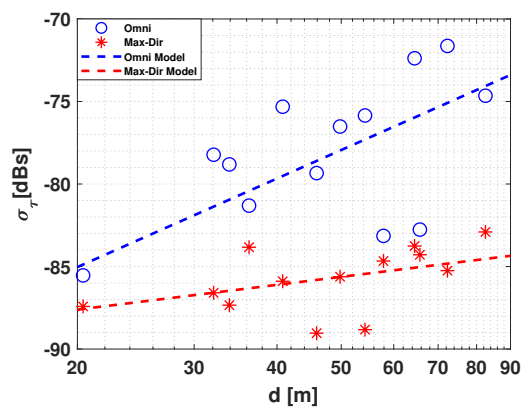}
		\caption{Linear modeling of $\sigma_\tau$ with weighting.}
		\vspace*{0mm}
		\label{fig:RMSDS-LOS-LF}%
	\end{subfigure}
	\caption{Modeling of delay spread for LoS cases.}%
	\vspace{-0 mm}
	\label{fig:RMSDS-LOS}%
	\vspace{-5mm}
\end{figure*}

The next parameter to evaluate is the RMSDS. In the LoS case, we expect lower values for the max-dir due to the spatial filtering. Similarly, an increase in the RMSDS with increasing distance between the Tx and Rx is expected, due to a large number and difference in runlength of the MPCs. Fig. \ref{fig:RMSDS-LOS}a shows the probability density function of the RMSDS. It is plotted on a logarithmic scale, i.e., dBs, as is common in particular in 3GPP. This representation also allows to easily see the excellent fit of a lognormal distribution to the measurement results. The variance of the max-dir points is approximately $62\%$ the value of the omni-directional case. 

Fig. \ref{fig:RMSDS-LOS}b shows the RMSDS as a function of distance and the linear regression, showing an increase with distance, as anticipated (and also in agreement with experimental results at lower frequencies).  It is also observed that for all measurement points the max-dir values are smaller than the omni-directional. 

\begin{figure*}[t!]
	\centering
	\begin{subfigure}[b]{0.45\textwidth}
		\centering
		%\hspace{7mm}
		\includegraphics[width=1\columnwidth]{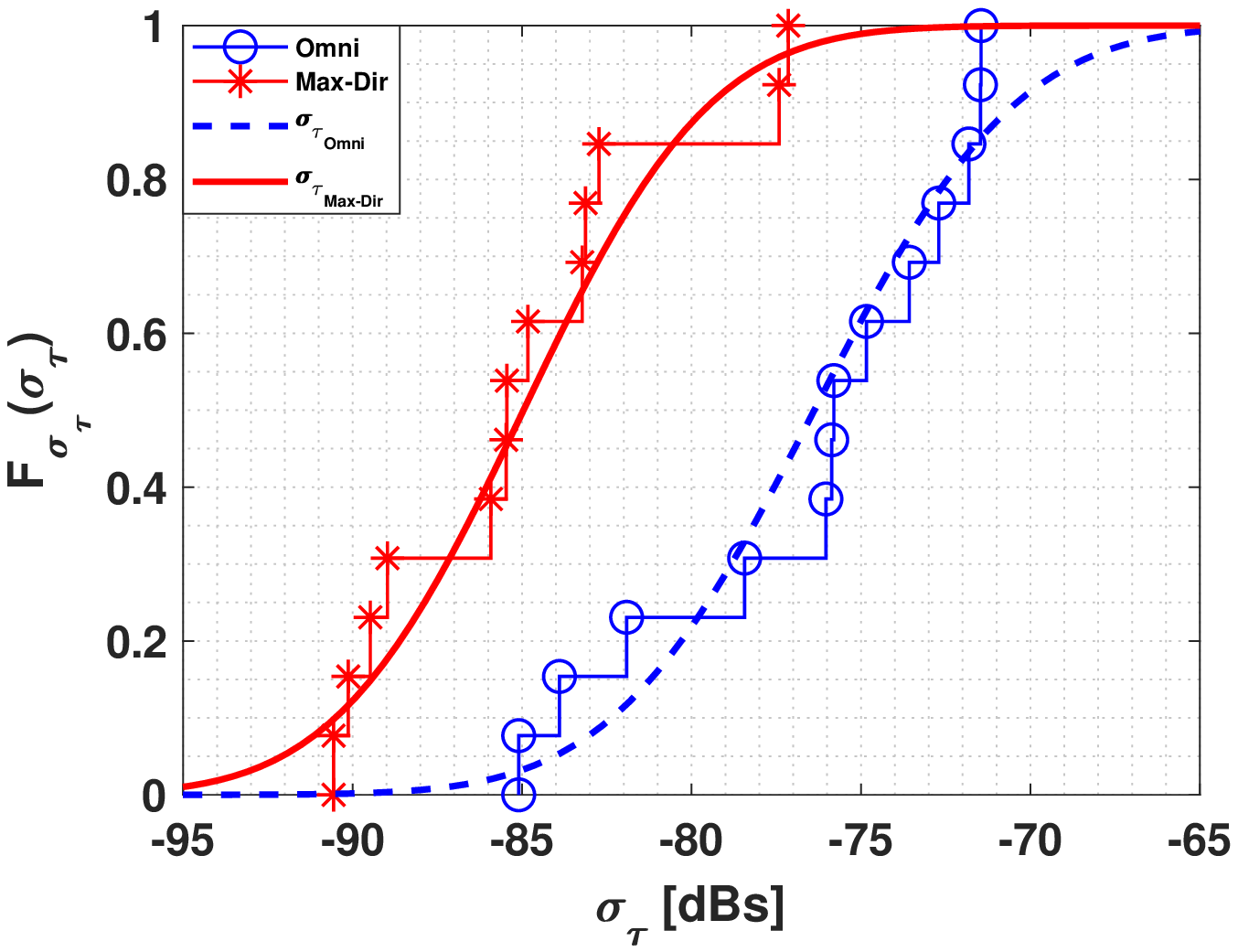}
		\caption{CDF}
		\vspace*{0mm}
		\label{fig:RMSDS-NLOS-CDF}%
	\end{subfigure}
	\begin{subfigure}[b]{0.45\textwidth}
		\centering
		%\hspace{7mm}
		\includegraphics[width=1\columnwidth]{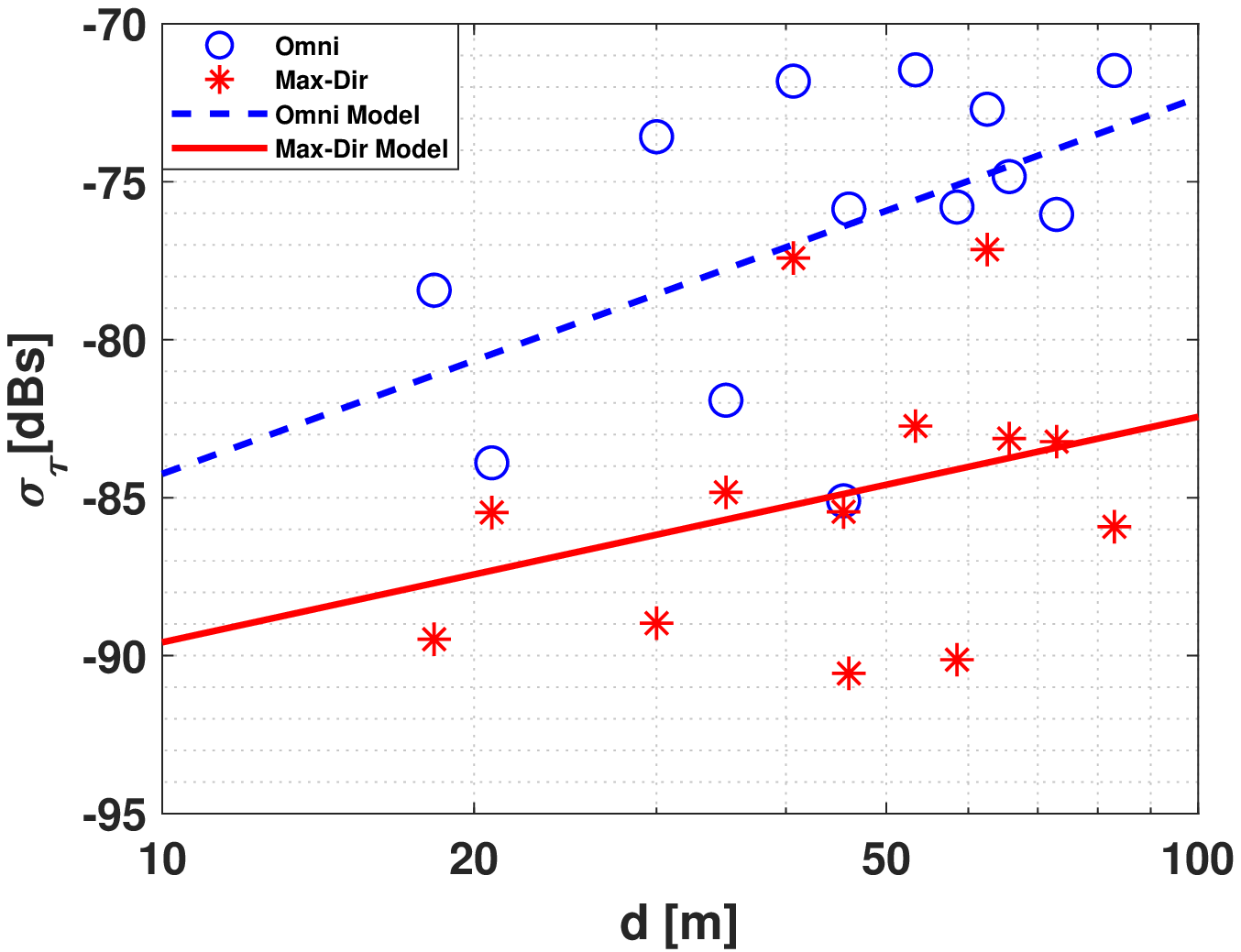}
		\caption{Linear fitting with $log_{10}(d)$ weighting.}
		\vspace*{0mm}
		\label{fig:RMSDS-NLOS-LF}%
	\end{subfigure}
	\caption{Modeling of delay spread for NLoS points.}%
	\vspace{-0 mm}
	\label{fig:RMSDS-NLOS}%
	\vspace{-5mm}
\end{figure*}

Fig. \ref{fig:RMSDS-NLOS} shows the RMSDS analysis for the NLoS case. It is observed that the CDFs have a different slope ($\beta_{omni}^{NLoS}=11.91, \beta_{max-dir}^{NLoS}=7.14$). This behavior can be related to the "street-canyon" scenarios of Routes One, Four, and Six. The waveguiding effect allows a concentration of the power and MPCs in a small set of directions, so the max-dir PDPs have low number of MPCs that are concentrated in smaller range of delay bins. A special case of the "waveguiding" effect is the point Tx5-Rx23 ($d=45.5m$), in which the $\sigma_\tau$ values for the omni-directional and max-dir cases are almost equal. A summary of the estimated regression parameters and the statistical analysis are shown in Tables \ref{tab:linear-model-RMSDS}, \ref{tab:RMSDS_CDF}.  

\begin{table*}[t!]
	\centering
	\caption{Linear model parameters for $\sigma_{\tau}$ with $95\%$ confidence interval.}
	\label{tab:linear-model-RMSDS}
	{%
		\begin{tabular}{|c|c|c|c|c|c|c|c|}
			\hline
			\multicolumn{1}{|c|}{\multirow{2}{*}{\textbf{Parameter}}} & \multicolumn{6}{c|}{\textbf{Linear model parameters estimated with 95\% CI}} \\ \cline{2-7} 
			\multicolumn{1}{|l|}{} & \multicolumn{1}{c|}{$\alpha$} & \multicolumn{1}{c|}{$\alpha_{min,95\%}$} & \multicolumn{1}{c|}{$\alpha_{max,95\%}$} & \multicolumn{1}{c|}{$\beta$} & \multicolumn{1}{c|}{$\beta_{min,95\%}$} & \multicolumn{1}{c|}{$\beta_{max,95\%}$} \\ \hline \hline
			$\sigma_{\tau_{omni}}^{LoS}$ & -108.22 & -122.6 & -93.83 & 17.82 & 8.76 & 26.88 \\ \hline
			$\sigma_{\tau_{max-dir}}^{LoS}$ & -94.11 & -100.92 & -87.29 & 4.99 & 0.7 & 9.28 \\ \hline
			$\sigma_{\tau_{omni}}^{NLoS}$ & -96.16 & -114.09 & -78.22 & 11.91 & 0.57 & 23.26 \\ \hline
			$\sigma_{\tau_{max-dir}}^{NLoS}$ & -96.71 & -113.8 & -79.63 & 7.14 & -3.67 & 17.95 \\ \hline
		\end{tabular}%
	}
\end{table*}

\begin{table*}[t!]
	\centering
	\caption{Statistical model parameters for $\sigma_{\tau}$ with $95\%$ confidence interval.}
	\label{tab:RMSDS_CDF}
	{%
		\begin{tabular}{|c|c|c|c|c|c|c|}
			\hline
			\multicolumn{1}{|c|}{\multirow{2}{*}{\textbf{Parameter}}} & \multicolumn{6}{c|}{\textbf{Statistical model parameters estimated with 95\% CI}} \\ \cline{2-7} 
			\multicolumn{1}{|l|}{} & \multicolumn{1}{c|}{$\mu$} & \multicolumn{1}{c|}{$\mu_{min,95\%}$} & \multicolumn{1}{c|}{$\mu_{max,95\%}$} & \multicolumn{1}{c|}{$\sigma$} & \multicolumn{1}{c|}{$\sigma_{min,95\%}$} & \multicolumn{1}{c|}{$\sigma_{max,95\%}$} \\ \hline \hline
			$\sigma_{\tau_{omni}}^{LoS}$ & -78.11 & -80.68 & -75.55 & 4.25 & 3.05 & 7.01 \\ \hline
			$\sigma_{\tau_{max-dir}}^{LoS}$ & -85.8 & -86.98 & -84.62 & 1.95 & 1.4 & 3.22 \\ \hline
			$\sigma_{\tau_{omni}}^{NLoS}$ & -76.38 & -79.2 & -73.56 & 4.66 & 3.34 & 7.7 \\ \hline
			$\sigma_{\tau_{max-dir}}^{NLoS}$ & -84.96 & -87.58 & -82.34 & 4.33 & 3.11 & 7.15 \\ \hline
		\end{tabular}%
	}
\end{table*}

\subsection{Angular spread}
The next parameter to analyze is the angular spread. In this case, the analysis is separated between the Tx and Rx end. As explained in Section II, the scan range for Tx and Rx are different,
so our conjecture is to observe a larger angular spread in the Rx side for both LoS and NLoS cases. Furthermore, the richer number of scattering objects at street level is expected to compound this effect. \\  

\begin{figure*}[t!]
	\centering
	\begin{subfigure}[b]{0.45\textwidth}
		\centering
		%\hspace{7mm}
		\includegraphics[width=1\columnwidth]{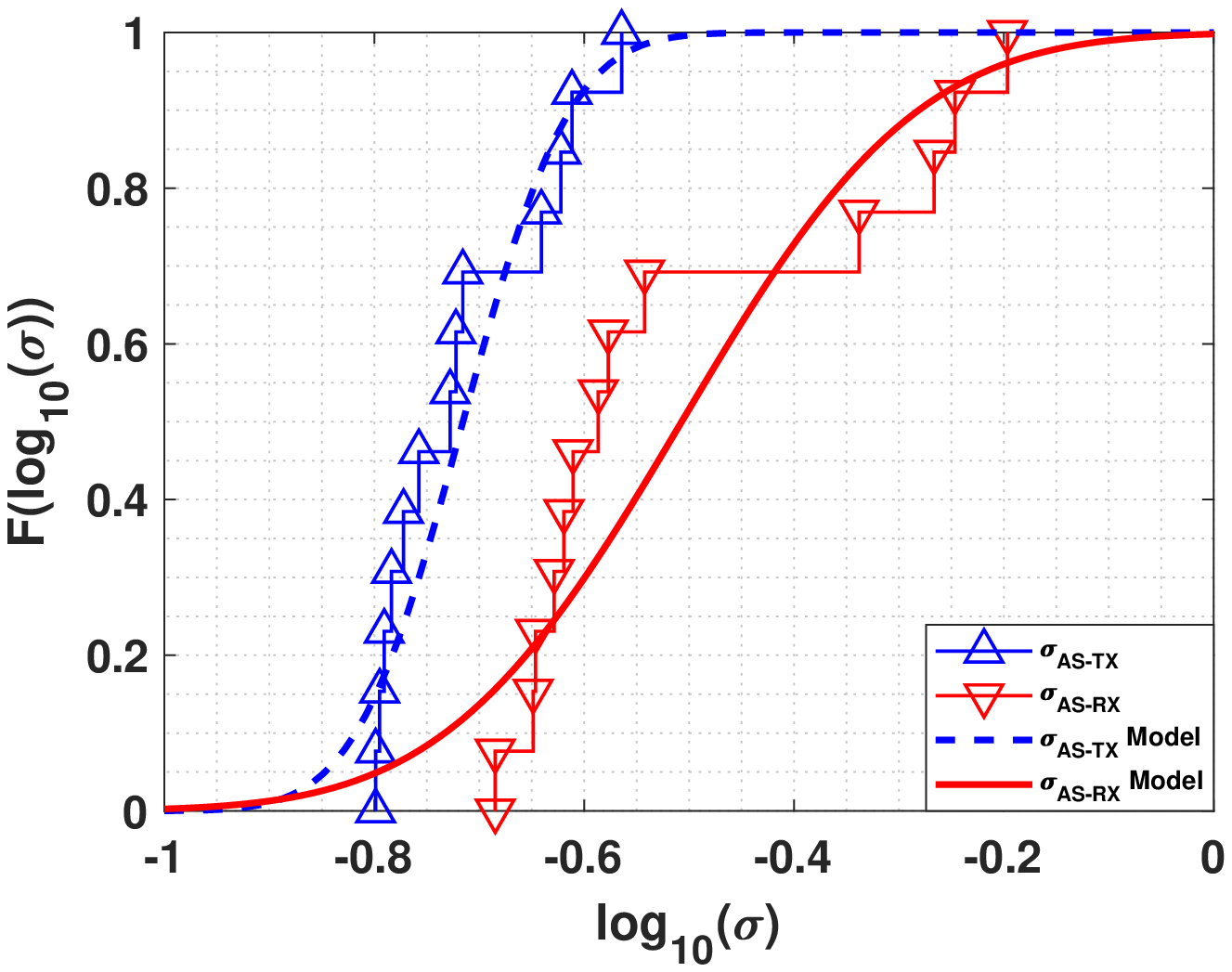}
		\caption{LoS case.}
		\vspace*{0mm}
		\label{fig:AS_CDF_LOS}%
	\end{subfigure}
	\begin{subfigure}[b]{0.45\textwidth}
		\centering
		%\hspace{7mm}
		\includegraphics[width=1\columnwidth]{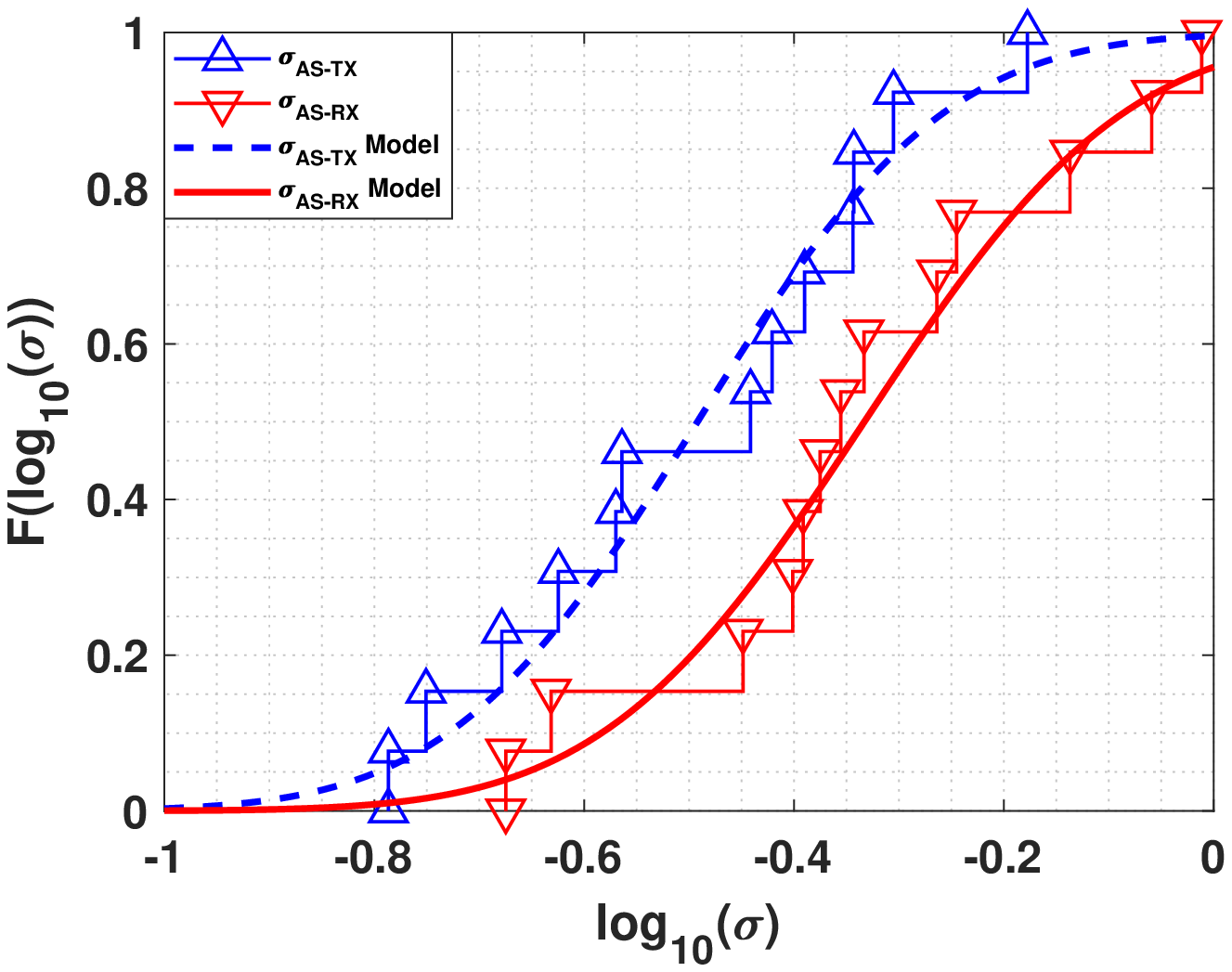}
		\caption{NLoS case.}
		\vspace*{0mm}
		\label{fig:AS_CDF_NLOS}%
	\end{subfigure}
	\caption{Modeling of $\sigma^\circ$ for all points.}%
	\vspace{-0 mm}
	\label{fig:AS-CDF}%
	\vspace{-5mm}
\end{figure*}

Fig. \ref{fig:AS-CDF} shows the CDF for LoS and NLoS cases. In both cases, the data confirm our hypothesis. For example, in the LoS case the Tx points show a smaller spread compared to the Rx ($\sigma^\circ_{NLoS} Tx < \sigma^\circ_{NLoS} Rx$). This result is related to the fact that reflected MPCs are reflected in the vicinity of the Rx, and are ''seen" by the Tx under angles similar to that of the LoS. On the other hand, the NLoS points show AS points with a similar spread (i.e. $\sigma^\circ_{NLoS} Tx \approx \sigma^\circ_{NLoS} Rx$). A possible cause for this behavior is the waveguiding in the "street canyon" environments, which concentrates the MPCs in a narrower angular range. A summary of the estimated statistical parameters with their $95\%$ confidence interval is shown in Table \ref{tab:stat-model-AS}. 

\begin{table*}[t!]
	\centering
	\caption{Statistical model parameters for $\sigma^\circ$ with $95\%$ confidence interval.}
	\label{tab:stat-model-AS}
	{%
		\begin{tabular}{|c|c|c|c|c|c|c|}
			\hline
			\multicolumn{1}{|c|}{\multirow{2}{*}{\textbf{Parameter}}} & \multicolumn{6}{c|}{\textbf{Statistical model parameters estimated with 95\% CI}} \\ \cline{2-7} 
			\multicolumn{1}{|l|}{} & \multicolumn{1}{c|}{$\mu$} & \multicolumn{1}{c|}{$\mu_{min,95\%}$} & \multicolumn{1}{c|}{$\mu_{max,95\%}$} & \multicolumn{1}{c|}{$\sigma$} & \multicolumn{1}{c|}{$\sigma_{min,95\%}$} & \multicolumn{1}{c|}{$\sigma_{max,95\%}$} \\ \hline \hline
			$\sigma^\circ_{LoS} Tx$ & -0.72 & -0.77 & -0.67 & 0.08 & 0.06 & 0.13 \\ \hline
			$\sigma^\circ_{LoS} Rx$ & -0.51 & -0.62 & -0.4 & 0.18 & 0.13 & 0.3 \\ \hline
			$\sigma^\circ_{NLoS} Tx$ & -0.49 & -0.6 & -0.38 & 0.18 & 0.13 & 0.3\\ \hline
			$\sigma^\circ_{NLoS} Rx$ & -0.33 & -0.45 & -0.21 & 0.19 & 0.14 & 0.32\\ \hline
		\end{tabular}%
	}
\end{table*}

\subsection{Power distribution of MPCs}

\begin{figure*}[t!]
	\centering
	\begin{subfigure}[b]{0.45\textwidth}
		\centering
		%\hspace{7mm}
		\includegraphics[width=1\columnwidth]{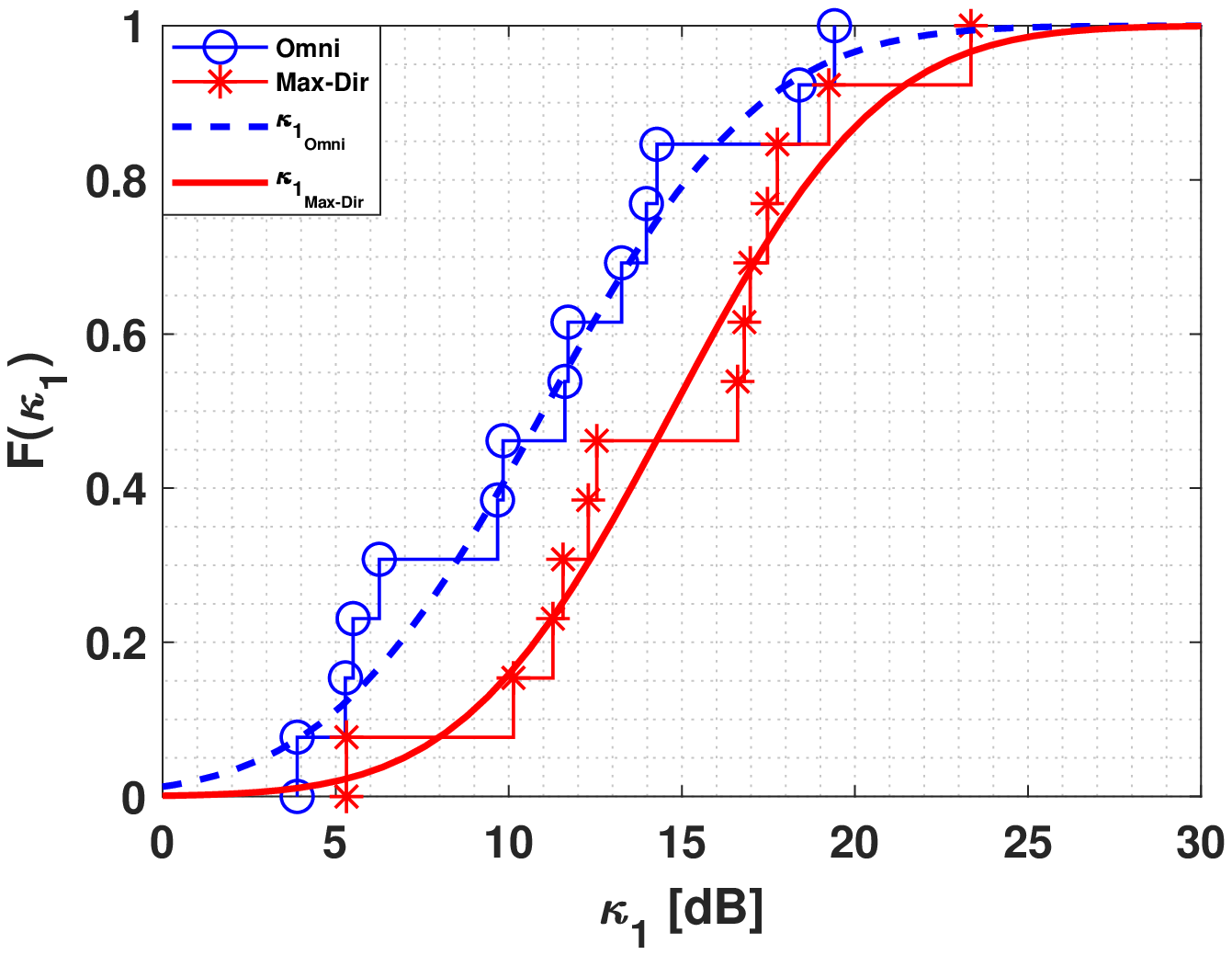}
		\caption{CDF.}
		\vspace*{0mm}
		\label{fig:k1_CDF}%

	\end{subfigure}
	\begin{subfigure}[b]{0.45\textwidth}
		\centering
		%\hspace{7mm}
		\includegraphics[width=1\columnwidth]{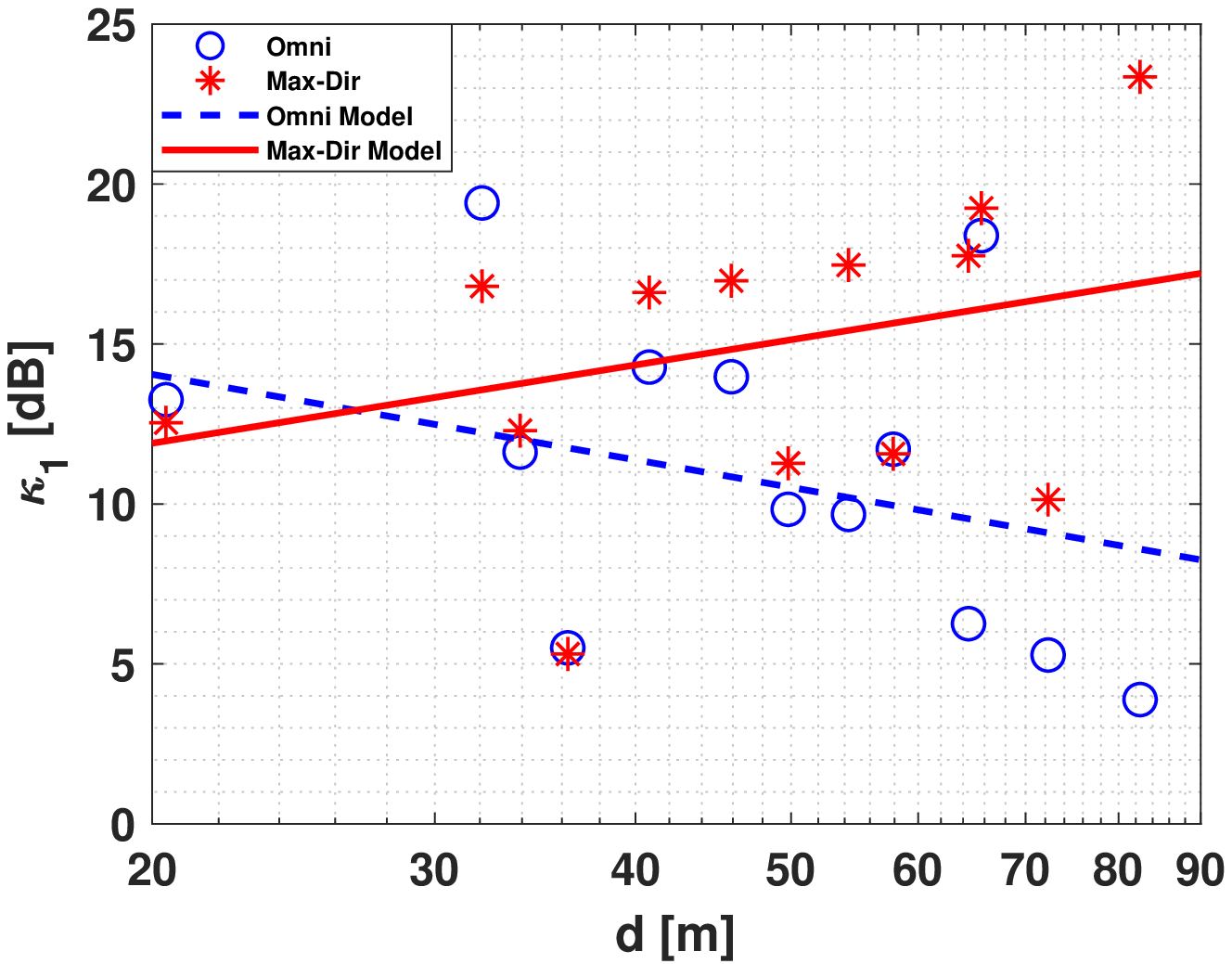}
		\caption{Linear fitting with $log_{10}(d)$ weighting.}
		\vspace*{0mm}
		\label{fig:k1_LS}%
	\end{subfigure}
	\caption{Modeling of $\kappa_1$ for LoS points.}%
	\vspace{-0 mm}
	\label{fig:k1-LOS}%
	\vspace{-5mm}
\end{figure*}

The final parameter estimated is the $\kappa_1$. Our hypothesis is to observe larger values of $\kappa_1$ in max-dir cases compared the omni-directional ones. Fig. \ref{fig:k1-LOS} shows the estimated values for the LoS case. As can be observed in Fig. \ref{fig:k1_CDF} the LoS points for the omni-directional case have a similar spread compared to the max-dir cases, but significantly smaller mean. Fig. \ref{fig:k1_LS} shows the regression analysis of the power distribution. The observed range oscillates between 4 and 23 dB. As observed in the plot, $\kappa_1$ for the max-dir grows as the distance increases and for the omni-directional case shows a decreasing trend. The filtering effect of the antenna decreases the number of MPCs received by the MPC. As the distance increases, additional MPCs (coming from reflections) suffer from further attenuation and only those in the LoS directions are boosted by the antenna gain. On the other hand, in the omni-directional case, the value of $\kappa_1$ decreases because as the distance increases more MPCs will be collected from different direction apart from the LoS\footnote{An unusual behavior is observed in point Tx1-Rx6 where ($\kappa_1^{omni} > \kappa_1^{max-dir}$), though the difference is small. We conjecture that this is caused by imperfections in the calibration procedure and the generation of omni-directional PDPs from the directional PDPs.}. A summary of the parameters see Tables  \ref{tab:linear-model-kappa}, \ref{tab:stat-model-kappa}. 

In the NLoS case, we observed the values with a range from -10 to 22 dB. This high variability  can be related to the multiple points in "street canyon" scenarios (Routes One, Five, and Six). The "street canyon"  filters/concentrates the MPCs arriving at the Rx. Furthermore, $\kappa_1$ is reduced when the distance increases, both for the omni- and the max-Dir case. Similarly to the RMSDS analysis, the points Tx5-Rx23 and Tx6-Rx24 shows a different behavior ($\kappa_{1_{omni}}^{NLoS}>\kappa_{1_{max-dir}}^{NLoS}$). This is related to the fact that the strongest MPC angle is between two azimuthal captures, which produces this unusual behavior. More details about the regression analysis and statistical modeling and estimation are shown in Tables \ref{tab:linear-model-kappa},\ref{tab:stat-model-kappa}.

\begin{figure*}[t!]
	\centering
	\begin{subfigure}[b]{0.45\textwidth}
		\centering
		%\hspace{7mm}
		\includegraphics[width=1\columnwidth]{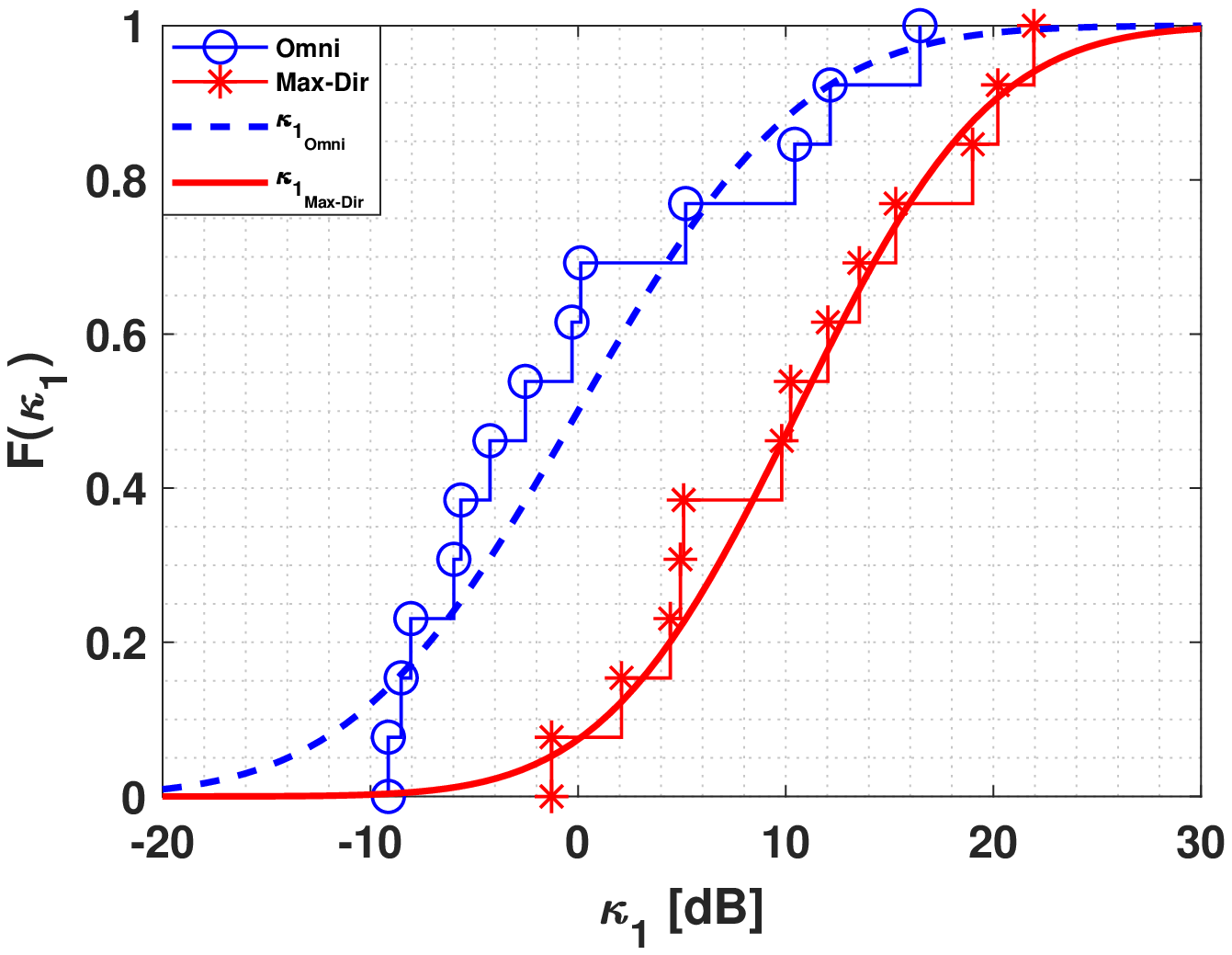}
		\caption{CDF.}
		\vspace*{0mm}
		\label{fig:k1_CDF_NLOS}%
	\end{subfigure}
	\begin{subfigure}[b]{0.45\textwidth}
		\centering
		%\hspace{7mm}
		\includegraphics[width=1\columnwidth]{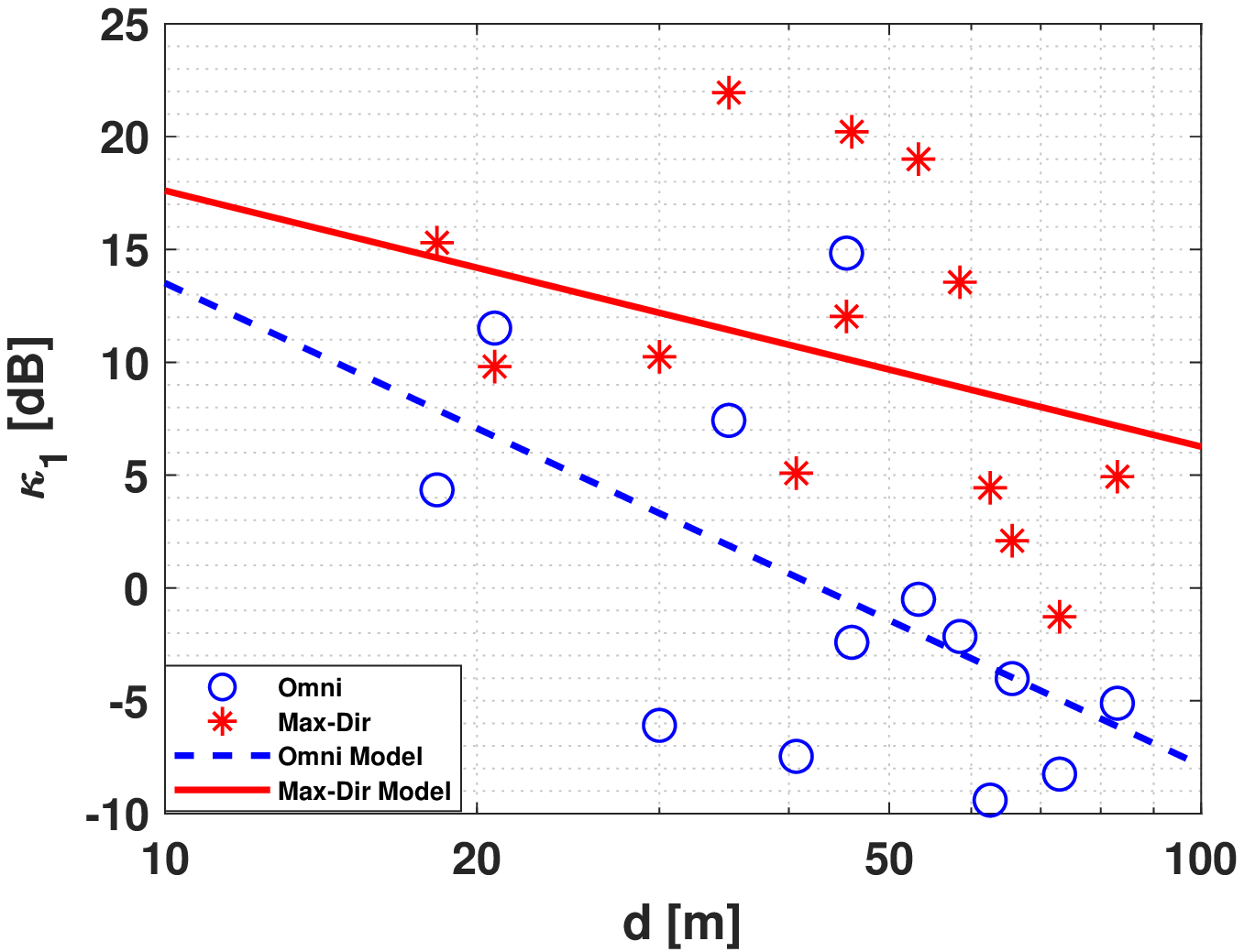}
		\caption{Linear fitting with $log_{10}(d)$ weighting.}
		\vspace*{0mm}
		\label{fig:k1_LS_NLOS}%
	\end{subfigure}
	\caption{Modeling of $\kappa_1$ for NLoS points.}%
	\vspace{-0 mm}
	\label{fig:k1_NLOS}%
	\vspace{-5mm}
\end{figure*}

\begin{table*}[t!]
	\centering
	\caption{Linear model parameters for $\kappa_1$ with $95\%$ confidence interval.}
	\label{tab:linear-model-kappa}
	{%
		\begin{tabular}{|c|c|c|c|c|c|c|c|}
			\hline
			\multicolumn{1}{|c|}{\multirow{2}{*}{\textbf{Parameter}}} & \multicolumn{6}{c|}{\textbf{Linear model parameters estimated with 95\% CI}} \\ \cline{2-7} 
			\multicolumn{1}{|l|}{} & \multicolumn{1}{c|}{$\alpha$} & \multicolumn{1}{c|}{$\alpha_{min,95\%}$} & \multicolumn{1}{c|}{$\alpha_{max,95\%}$} & \multicolumn{1}{c|}{$\beta$} & \multicolumn{1}{c|}{$\beta_{min,95\%}$} & \multicolumn{1}{c|}{$\beta_{max,95\%}$} \\ \hline \hline
			$\kappa_{1_{omni}}^{LoS}$ & 25.59 & 7.49 & 43.7 & -8.87 & -20.27 & 2.53\\ \hline
			$\kappa_{1_{max-dir}}^{LoS}$ & 1.32 & -16.25 & 18.89 & 8.13 & -2.94 & 19.19 \\ \hline
			$\kappa_{1_{omni}}^{NLoS}$ & 38.54 & 8.12 & 68.96 & -23.29 & -42.54 & -4.05\\ \hline
			$\kappa_{1_{max-dir}}^{NLoS}$ & 28.95 & 1.39 & 56.52 & -11.35 & -28.78 & 6.09 \\ \hline
		\end{tabular}%
	}
\end{table*}

\begin{table*}[t!]
	\centering
	\caption{Statistical model parameters for $\kappa_1$ with $95\%$ confidence interval.}
	\label{tab:stat-model-kappa}
	{%
		\begin{tabular}{|c|c|c|c|c|c|c|}
			\hline
			\multicolumn{1}{|c|}{\multirow{2}{*}{\textbf{Parameter}}} & \multicolumn{6}{c|}{\textbf{Statistical model parameters estimated with 95\% CI}} \\ \cline{2-7} 
			\multicolumn{1}{|c|}{} & \multicolumn{1}{c|}{$\mu$} & \multicolumn{1}{c|}{$\mu_{min,95\%}$} & \multicolumn{1}{c|}{$\mu_{max,95\%}$} & \multicolumn{1}{c|}{$\sigma$} & \multicolumn{1}{c|}{$\sigma_{min,95\%}$} & \multicolumn{1}{c|}{$\sigma_{max,95\%}$} \\ \hline \hline
			$\kappa_{1_{omni}}^{LoS}$ & 11.01 & 8.03 & 13.98 & 4.92 & 3.53 & 8.12 \\ \hline
			$\kappa_{1_{max-dir}}^{LoS}$ & 14.72 & 11.87 & 17.57 & 4.72 & 3.38 & 7.78 \\ \hline
			$\kappa_{1_{omni}}^{NLoS}$ & 0 & -5.14 & 5.13 & 8.5 & 6.1 & 14.03 \\ \hline
			$\kappa_{1_{max-dir}}^{NLoS}$ & 10.57 & 6.15 & 14.98 & 7.3 & 5.23 & 12.05 \\ \hline
		\end{tabular}%
	}
\end{table*}

\subsection{Summary of results}

In this section, a summary of the estimated parameter for a systems design or channel simulation are shown in Tables \ref{tab:linear-model-summary}, \ref{tab:stat-model-summary}. Table \ref{tab:linear-model-summary} shows the regression analysis (i.e. linear modeling) for the distance dependence of the parameters for both LoS and NLoS cases. Table \ref{tab:stat-model-summary} shows the estimated parameters for the statistical fits/modeling carried out in this analysis for both LoS and NLoS cases. Please note that the presented statistical results are valid for the ranges of distances we measured over ($\approx 20-\approx 85$ m). 

It is important to note that the parameters obtained in the analysis are directly related to the number of points and the selection of measurement locations. In other words, this analysis is impacted by the fact that the measurement locations were chosen such that reasonable Rx power could be anticipated. An analysis of outage probability should consider a ''blind" selection of points, e.g., on a regular grid, that would allow an assessment of the percentage of points that cannot sustain communications at a given sensitivity level.  Also other parameters, which might be correlated to the received power, might conceivably be influenced by the selection of the points. The results in this paper should thus be interpreted as ''conditioned on the existence of reasonable Rx power". 

Furthermore, while in the current campaign more than 100,000 transfer functions were measured, the number of measured {\em location pairs} is still somewhat limited. Hence, this model is based on a relatively small number of points to provide an initial channel model to give a realistic analysis and model for system design. A larger number of measurement locations will obviously increase the number of measurement locations and increase the validity of the analysis. However, the time required to perform the current campaign was quite significant (several months), and it is among the largest double-directional campaigns ever performed in the THz regime (for any type of environment). Future measurements will be added to improve the model further.      

\begin{table*}[t!]
	\centering
	\caption{Linear model parameters summary.}
	\label{tab:linear-model-summary}
	{%
		\begin{tabular}{|c|c|c|}
			\hline
			\multicolumn{1}{|c|}{\textbf{Parameter}} & \multicolumn{1}{c|}{$\alpha$} & \multicolumn{1}{c|}{$\beta$} \\ \hline \hline
			$PL_{omni}^{LoS}$ & 72.88 & 1.93 \\ \hline
			$PL_{max-dir}^{LoS}$ & 77.33 & 1.88 \\ \hline
			$PL_{omni}^{LoS} OLS$ &  75.02 & 1.8 \\ \hline
			$PL_{max-dir}^{LoS} OLS$ & 77.06 & 1.89 \\ \hline
			$\sigma_{\tau_{omni}}^{LoS}$ & -108.22 & 17.82 \\ \hline
			$\sigma_{\tau_{max-dir}}^{LoS}$ & -94.11 & 4.99 \\ \hline
			$\kappa_{1_{omni}}^{LoS}$ & 25.59 & -8.87  \\ \hline
			$\kappa_{1_{max-dir}}^{LoS}$ & 1.32 & 8.13 \\ \hline
			$PL_{omni}^{NLoS}$ & 91.28 & 1.76 \\ \hline
			$PL_{max-dir}^{NLoS}$ & 84.54 & 2.57 \\ \hline
			$PL_{omni}^{NLoS} OLS$ & 86.81 & 2.03 \\ \hline
			$PL_{max-dir}^{NLoS} OLS$ & 82.91 & 2.68 \\ \hline
			$\sigma_{\tau_{omni}}^{NLoS}$ & -96.16 & 11.91\\ \hline
			$\sigma_{\tau_{max-dir}}^{NLoS}$ & -96.71 & 7.14\\ \hline
			$\kappa_{1_{omni}}^{NLoS}$ & 38.54 & -23.29\\ \hline
			$\kappa_{1_{max-dir}}^{NLoS}$ & 28.95 & -11.35 \\ \hline
		\end{tabular}%
	}
\end{table*}

\begin{table*}[t!]
	\centering
	\caption{Statistical model parameters summary.}
	\label{tab:stat-model-summary}
	{%
		\begin{tabular}{|c|c|c|}
			\hline
			\multicolumn{1}{|c|}{\textbf{Parameter}} & \multicolumn{1}{c|}{$\mu$} & \multicolumn{1}{c|}{$\sigma$}\\ \hline \hline
			$\epsilon_{omni}^{LoS}$ & 0.09 & 0.72 \\ \hline
			$\epsilon_{max-dir}^{LoS}$ & -0.01 & 0.8 \\ \hline
			$\epsilon_{omni}^{LoS} OLS$ & 0 & 0.69\\ \hline
			$\epsilon_{max-dir}^{LoS} OLS$ & 0 & 0.8\\ \hline
			$\sigma^{\circ}_{LoS} Tx$ & -0.72 & 0.08\\ \hline
			$\sigma^{\circ}_{LoS} Rx$ & -0.51 & 0.18\\ \hline
			$\sigma_{\tau_{omni}}^{LoS}$ & -78.11 & 4.25\\ \hline
			$\sigma_{\tau_{max-dir}}^{LoS}$ & -85.8 & 1.95 \\ \hline
			$\kappa_{1_{omni}}^{LoS}$ & 11.01 & 4.92 \\ \hline
			$\kappa_{1_{max-dir}}^{LoS}$ & 14.72 & 4.72 \\ \hline
			$\epsilon_{omni}^{NLoS}$ & 0.04 & 6.24 \\ \hline
			$\epsilon_{max-dir}^{NLoS}$ & 0.18 & 6.21\\ \hline
			$\epsilon_{omni}^{NLoS} OLS$ & 0 & 6.21 \\ \hline
			$\epsilon_{max-dir}^{NLoS} OLS$ & 0 & 7.89 \\ \hline
			$\sigma^{\circ}_{NLoS} Tx$ & -0.49 & 0.18 \\ \hline
			$\sigma^{\circ}_{NLoS} Rx$ & -0.33 & 0.19 \\ \hline
			$\sigma_{\tau_{omni}}^{NLoS}$ & -76.38 & 4.66 \\ \hline
			$\sigma_{\tau_{max-dir}}^{NLoS}$ & -84.96 & 4.33 \\ \hline
			$\kappa_{1_{omni}}^{NLoS}$ & 0 & 8.5 \\ \hline
			$\kappa_{1_{max-dir}}^{NLoS}$ & 10.57 & 7.37\\ \hline
		\end{tabular}%
	}
\end{table*}

\section{Conclusions}
In this paper, we presented the results of the first extensive wideband, double-directional THz outdoor channel measurements for microcell scenarios with Tx heights of more than 10 m above the ground. We provide an overview of the measurement methodology and environments, as well as the signal processing to extract parameters characterizing the channels. Most importantly, we provided a parameterized statistical description of our measurement results that can be used to assess THz systems. The key parameters discussed in the current paper include path loss, shadowing, angular spread, delay spread and MPC power distribution.
%We note that our current results are from the perspective of the current measurement system and the corresponding antennas. 
These results are an important step towards drawing some important first conclusions about the implications on system design and deployment in the THz regime.

\section*{Acknowledgment}
%The authors would like to thank the Semiconductor Research Corporation (SRC), whose ComSenTer project partially funded this work. 
Helpful discussions with Sundeep Rangan, Mark Rodwell and Zihang Cheng are gratefully acknowledged. 
%The work of AFM was also supported by the National Science Foundation. The work of JGP was supported by the Foreign Fulbright Ecuador SENESCYT Program.

% Generated by IEEEtran.bst, version: 1.14 (2015/08/26)

\end{document}